\documentclass[10pt, preprint2]{emulateapj}
\usepackage{verbatim}
\usepackage{amsmath}
\usepackage[usenames, dvipsnames]{color}
\usepackage[pagebackref]{hyperref}
\usepackage{changepage}

\newcommand{\target}{TIC 220568520}

\newcommand{\rsun}{\ensuremath{R_\sun}}
\newcommand{\msun}{\ensuremath{M_\sun}}
\newcommand{\lsun}{\ensuremath{L_\sun}}

\newcommand{\rjup}{\ensuremath{R_{\rm Jup}}}
\newcommand{\mjup}{\ensuremath{M_{\rm Jup}}}
\newcommand{\teff}{\ensuremath{T_{\rm eff}}}

\newcommand{\logg}{\ensuremath{\log g}}
\newcommand{\feh}{[Fe/H]}
\newcommand{\vsini}{\ensuremath{V\sin(I)}}

\newcommand{\kms}{\ensuremath{\rm km\,s^{-1}}}
\newcommand{\ms}{\ensuremath{\rm m\,s^{-1}}}
\newcommand{\rstar}{\ensuremath{R_s}}
\newcommand{\mstar}{\ensuremath{M_s}}

\newcommand{\mic}{\ensuremath{\mu \rm m}}
\newcommand{\rpri}{\ensuremath{R_1}}
\newcommand{\mpri}{\ensuremath{M_1}}
\newcommand{\lpri}{\ensuremath{L_1}}
\newcommand{\rsec}{\ensuremath{R_2}}
\newcommand{\msec}{\ensuremath{M_2}}

\newcommand{\tess}{{\it TESS}}

\newcommand{\figr}[1]{Figure~\ref{fig:#1}}
\newcommand{\secr}[1]{Section~\ref{sec:#1}}

\newcommand{\tabr}[1]{\mbox{Table~\ref{tab:#1}}}

\newcommand{\gcmc}{\ensuremath{\rm g\,cm^{-3}}}

\shorttitle{}
\shortauthors{}

\begin{document}

\title{TOI 694 \MakeLowercase{b} and TIC 220568520 \MakeLowercase{b}: Two Low-mass Companions Near the Hydrogen Burning Mass Limit Orbiting Sun-like Stars}

\author{
Ismael Mireles\altaffilmark{1}, 
Avi Shporer\altaffilmark{1}, 
Nolan Grieves\altaffilmark{2},
George Zhou\altaffilmark{3},
Maximilian N.~G{\"u}nther\altaffilmark{1,A}, 
Rafael Brahm\altaffilmark{4,5},
Carl Ziegler\altaffilmark{6},
Keivan G.~Stassun\altaffilmark{7}, 
Chelsea X.~Huang\altaffilmark{1}, 
Louise Nielsen\altaffilmark{2},
Leonardo A. dos Santos\altaffilmark{2},
St\'{e}phane Udry\altaffilmark{2},
François Bouchy\altaffilmark{2},
Michael Ireland\altaffilmark{8},
Alexander Wallace\altaffilmark{8},
Paula Sarkis\altaffilmark{9},
Thomas Henning\altaffilmark{9},
Andr{\'e}s Jor{d\'a}n\altaffilmark{4,5},
Nicholas Law\altaffilmark{10},
Andrew W.~Mann\altaffilmark{10}, 
Leonardo A.~Paredes\altaffilmark{11},
Hodari-Sadiki James\altaffilmark{11},
Wei-Chun Jao\altaffilmark{11},
Todd J.~Henry\altaffilmark{12},
R. Paul Butler\altaffilmark{13}, 
Joseph E. Rodriguez\altaffilmark{14}, 
Liang Yu\altaffilmark{1}, 
Erin Flowers\altaffilmark{15, B}, 
George R.~Ricker\altaffilmark{1}, 
David W.~Latham\altaffilmark{3}, 
Roland Vanderspek\altaffilmark{1}, 
Sara Seager\altaffilmark{1, 16, 17}, 
Joshua N.~Winn\altaffilmark{15}, 
Jon M.~Jenkins\altaffilmark{18}, 
Gabor Furesz\altaffilmark{1},
Katharine Hesse\altaffilmark{19},
Elisa V.~Quintana\altaffilmark{20},
Mark E.~Rose\altaffilmark{18},
Jeffrey C.~Smith\altaffilmark{18,21}, 
Peter Tenenbaum\altaffilmark{18,21},
Michael Vezie\altaffilmark{16},
Daniel A.~Yahalomi\altaffilmark{3}, 
Zhuchang Zhan\altaffilmark{16} 
}

\altaffiltext{1}{Department of Physics and Kavli Institute for Astrophysics and Space Research, Massachusetts Institute of Technology, Cambridge, MA 02139, USA}
\altaffiltext{2}{Observatoire astronomique de l\'{}Universit\'{e} de Gen\`{e}ve, 51 Chemin des Maillettes, 1290 Versoix, Switzerland}
\altaffiltext{3}{Harvard-Smithsonian Center for Astrophysics, 60 Garden Street, Cambridge, MA 02138, USA}
\altaffiltext{4}{Facultad de Ingenier{\'i}a y Ciencias, Universidad Adolfo Ib{\'a}\~{n}ez, Av. Diagonal las Torres 2640, Pe\~{n}alol{\'e}n, Santiago, Chile}
\altaffiltext{5}{Millennium Institute for Astrophysics, Chile}
\altaffiltext{6}{Dunlap Institute for Astronomy and Astrophysics, University of Toronto, 50 St. George Street, Toronto, Ontario M5S 3H4, Canada}
\altaffiltext{7}{Vanderbilt University, Department of Physics \& Astronomy, 6301 Stevenson Center Lane, Nashville, TN 37235, USA}
\altaffiltext{8}{Research School of Astronomy and Astrophysics, Australian National University, Canberra, ACT 2611, Australia}
\altaffiltext{9}{Max-Planck-Institut f{\"u}r Astronomie, K{\"o}nigstuhl 17, 69117 Heidelberg, Germany}
\altaffiltext{10}{{Department of Physics and Astronomy, The University of North Carolina at Chapel Hill, Chapel Hill, NC 27599-3255, USA}}
\altaffiltext{11}{Physics and Astronomy Department, Georgia State University, Atlanta, GA 30302, USA}
\altaffiltext{12}{RECONS Institute, Chambersburg, PA, USA}
\altaffiltext{13}{Earth \&
Planets Laboratory, Carnegie Institution for Science, 5241 Broad Branch Road, NW, Washington, DC 20015, USA}
\altaffiltext{14}{Center for Astrophysics \textbar \ Harvard \& Smithsonian, 60 Garden St, Cambridge, MA 02138, USA}
\altaffiltext{15}{Department of Astrophysical Sciences, Princeton University, Princeton, NJ 08544, USA}
\altaffiltext{16}{Department of Earth, Atmospheric, and Planetary Sciences, Massachusetts Institute of Technology, Cambridge, MA 02139, USA}
\altaffiltext{17}{Department of Aeronautics and Astronautics, MIT, 77 Massachusetts Avenue, Cambridge, MA 02139, USA}
\altaffiltext{18}{NASA Ames Research Center, Moffett Field, CA 94035, USA}
\altaffiltext{19}{Department of Astronomy, Wesleyan University, Middletown, CT 06459, USA}
\altaffiltext{20}{NASA Goddard Space Flight Center, 8800 Greenbelt Road, Greenbelt, MD 20771, USA}
\altaffiltext{21}{SETI Institute, Mountain View, CA 94043, USA}

\altaffiltext{A}{Juan Carlos Torres Fellow}
\altaffiltext{B}{NSF Graduate Research Fellow}

\begin{abstract}

We report the discovery of TOI 694 b and TIC 220568520 b, two low-mass stellar companions in eccentric orbits around metal-rich Sun-like stars, first detected by the Transiting Exoplanet Survey Satellite (\tess). TOI 694 b has an orbital period of 48.05131$\pm$0.00019 days and eccentricity of 0.51946$\pm$0.00081, and we derive a mass of 89.0$\pm$5.3 \mjup\ (0.0849$\pm$0.0051 \msun) and radius of 1.111$\pm$0.017 \rjup\ (0.1142$\pm$0.0017 \rsun). TIC 220568520 b has an orbital period of 18.55769$\pm$0.00039 days and eccentricity of 0.0964$\pm$0.0032, and we derive a mass of 107.2$\pm$5.2 \mjup\ (0.1023$\pm$0.0050 \msun) and radius of 1.248$\pm$0.018 \rjup\ (0.1282$\pm$0.0019 \rsun). Both binary companions lie close to and above the Hydrogen burning mass threshold that separates brown dwarfs and the lowest mass stars, with TOI 694 b being 2-$\sigma$ above the canonical mass threshold of 0.075 \msun. The relatively long periods of the systems mean that the magnetic fields of the low-mass companions are not expected to inhibit convection and inflate the radius, which according to one leading theory is common in similar objects residing in short-period tidally-synchronized binary systems. Indeed we do not find radius inflation for these two objects when compared to theoretical isochrones. These two new objects add to the short but growing list of low-mass stars with well-measured masses and radii, and highlight the potential of the \tess\, mission for detecting such rare objects orbiting bright stars. 

\end{abstract}

\section{Introduction}
\label{sec:intro}

The stellar initial mass function (IMF) shows a maximum at or close to 0.1~\msun, at the bottom of the main sequence \citep[see e.g.,][and references therein]{chabrier03, bonnell07, andersen08, krumholz14}. That, combined with the increase in main sequence lifetime with decreasing stellar mass means that stars around 0.1~\msun\ are the most abundant stars in the galaxy. However, despite their abundance the precise measurement of their stellar properties, specifically radius and mass, is hindered by their low luminosity, leading to only a small number of objects at $\sim$0.1~\msun\ with precisely measured radius and mass. 

Precise measurements of those properties are desirable, for example, when low-mass stars are found to host planets, since the precision of the measured planet parameters depends on the precision of the host star parameters. The latter is especially relevant to low-mass stars since their small radius and mass provides an opportunity to detect smaller planets orbiting them compared to larger, Sun-like stars (this is commonly known as the ``M-dwarf opportunity", see e.g.~\citealt{gould2003, nutzman08}).

One way to detect such low-mass stars whose mass and radius can be precisely measured is through wide-field photometric transit surveys. The signal searched for in those surveys scales quadratically with the radius of the transiting object, and the radius of the smallest stars is the same as that of gas giant planets, resulting in similar signals. This was already exemplified by several discoveries \citep[e.g.,][]{pont2005, diaz2014, zhou2014_rv, chaturvedi2016, vonBoetticher2019}.

A comparison between the sizes and masses of low-mass stars to theoretical predictions has shown that the measured stellar radii tend to be larger than expected \citep[e.g.][]{ribas06, Torres:2010, kesseli2018} by about 5--10\%. The leading hypothesis for the inflated stellar radius is strong magnetic fields that inhibit convection, leading to decreased heat flow and in turn increased radius \citep{Chabrier2007}. The magnetic fields are strengthened by the relatively fast stellar rotation resulting from spin-orbit tidal synchronization in short orbital period systems \citep{mazeh2008}. Given that these systems are detected through photometric transit surveys it is not surprising they tend to have short periods (shorter than $\sim$10~days). 

Here we present the discovery of two longer-period eclipsing binary systems with low-mass companions close to 0.1~\msun\ detected by the Transiting Exoplanet Survey Satellite (\tess) mission:  TOI~694~b, with a period of 48.1 days, and TIC~220568520~b, with a period of 18.6 days. The long orbital periods mean that the companions are not expected to be tidally synchronized, so they are not expected to rotate rapidly and their magnetic fields are not expected to be strong enough to impact the stellar radius. We describe the \tess\ and ground-based observations in \secr{obs}, and the data analysis in \secr{data_analysis}. We discuss our results in \secr{dis} and conclude with a summary in \secr{sum}.


\section{Observations}
\label{sec:obs}

Basic photometric and astrometric information about the two targets studied here is given in \tabr{info}.

\subsection{\tess\ Photometry}
\label{sec:tess}

\begin{figure*}
\begin{center}
\includegraphics[width=\textwidth]{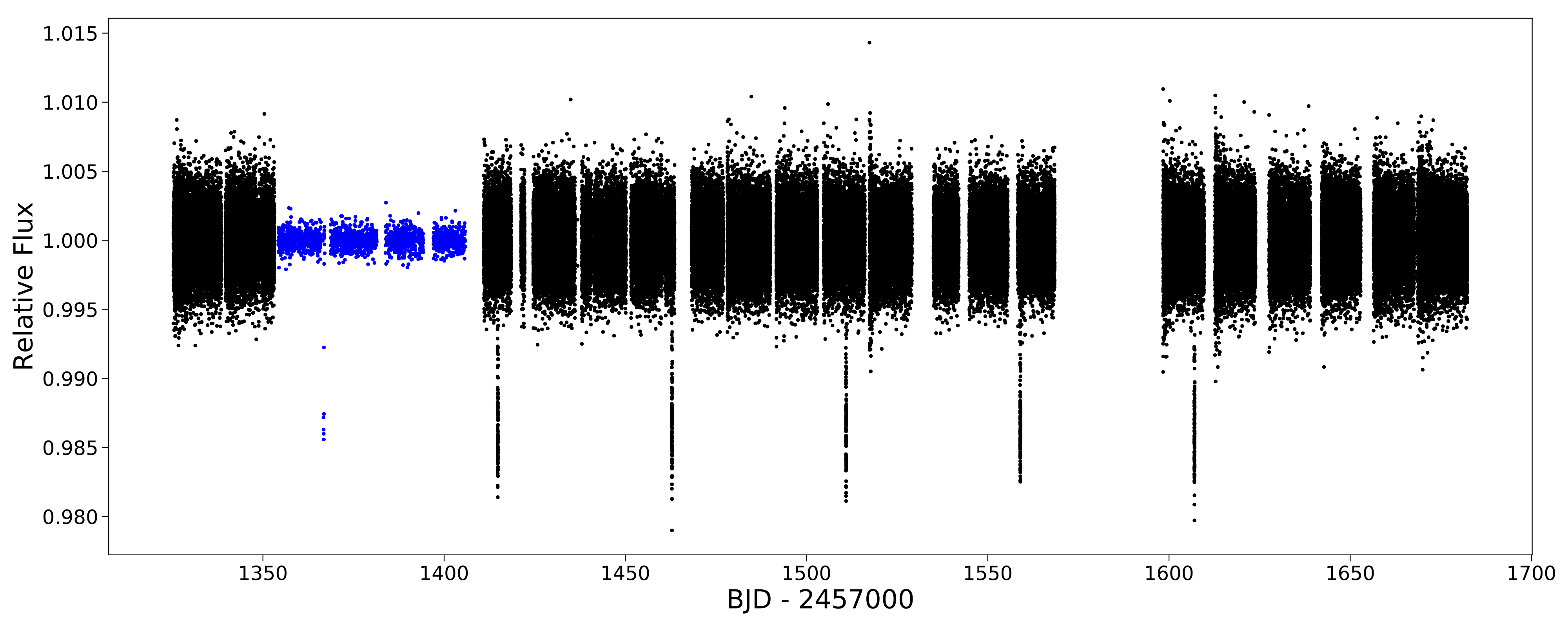}
\end{center}
\caption{The detrended SPOC 2-minute (black) and QLP 30-minute cadence (blue) \tess\ light curve of TOI 694. Six transits are clearly seen, 5 of which were observed with 2-minute cadences.}
\label{tess_lc_694}
\end{figure*}

\begin{figure*}
\includegraphics[width=\columnwidth]{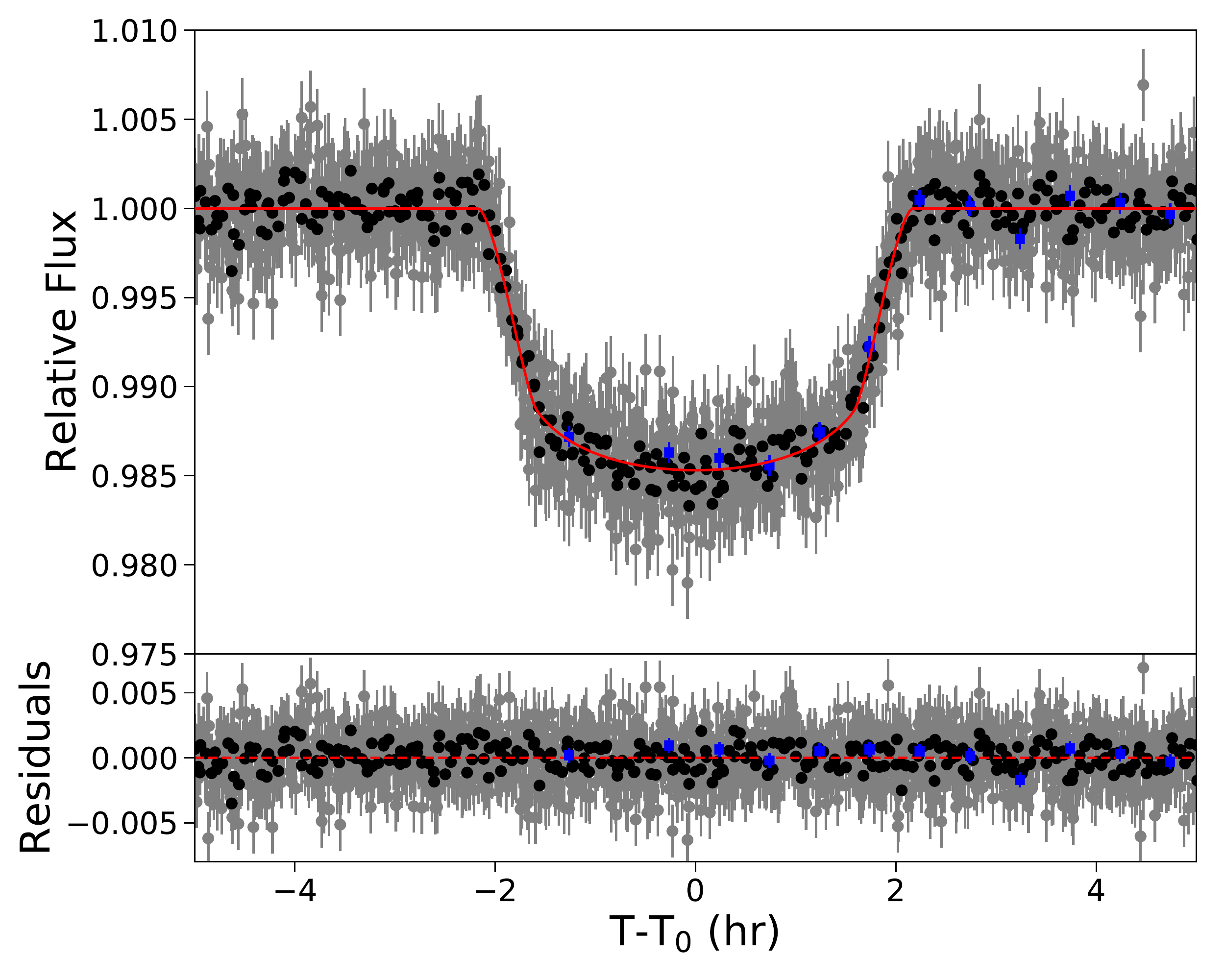}
\includegraphics[width=\columnwidth]{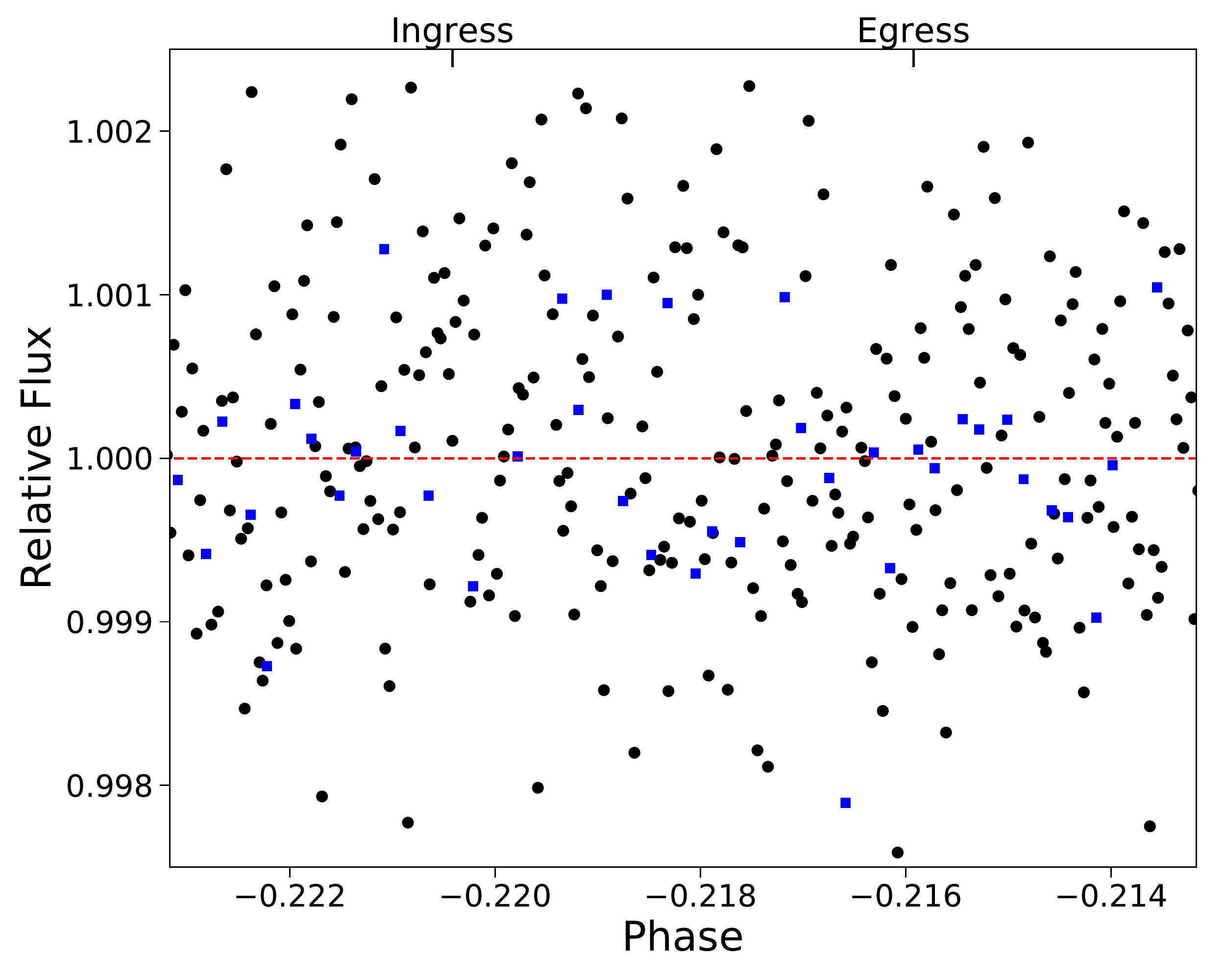}
\caption{Zoom-in view of the phased \tess\ 2-minute cadence (grey), binned 2-minute cadence (black), and 30-minute cadence (blue) light curve of the TOI 694 transit (left panel) and secondary eclipse (right panel). In the left panel, the red line is the fitted transit model, and the residuals (data - model) are shown at the bottom. 
In the right panel, the dashed red line is the median relative flux value out of transit. The error bars are not shown in order to improve visibility. The ingress and egress points are denoted at the top of the figure.}
\label{tess_lc_sec_694}
\end{figure*}

TOI~694 (TIC~55383975; \textit{V} = 11.963 mag) was observed by \tess\ Camera 4 during a total of 12 Sectors. It was observed in 2-minute cadence on Sector 1 (UT 2018 July 25 to UT 2018 August 22), Sectors 4 through 9 (UT 2018 October 18 to UT 2019 March 26), and Sectors 11 through 13 (UT 2019 April 22 to UT 2019 July 18). The target was within the field of view also in Sectors 2 and 3 (UT 2018 August 22 to UT 2018 October 18), but since it was not included in the pixel stamps observed with 2 minute exposures during those sectors we use the Full Frame Image (FFI) observations with a cadence of 30 minutes. \tess\ observations were interrupted between each of the 13.7 day long orbits of the satellite when data were downloaded to Earth. The 2-minute data were processed by the Science Processing Operations Center (SPOC; \citealt{jenkins2016}) pipeline which produced two light curves per sector called Simple Aperture Photometry (SAP) and Presearch Data Conditioning Simple Aperture Photometry (PDCSAP; \citealt{pdcsap_2012}; \citealt{pdcsap_2012_st}; \citealt{pdcsap_2014}) and identified 1.5\% deep transit-like dips every 48.1 days. The FFIs were processed by the Quick Look Pipeline (QLP; Huang et al.~in prep.). The \tess\ light curves are shown in Figures \ref{tess_lc_694} and \ref{tess_lc_sec_694}. 

\begin{figure*}
\begin{center}
\includegraphics[width=\textwidth]{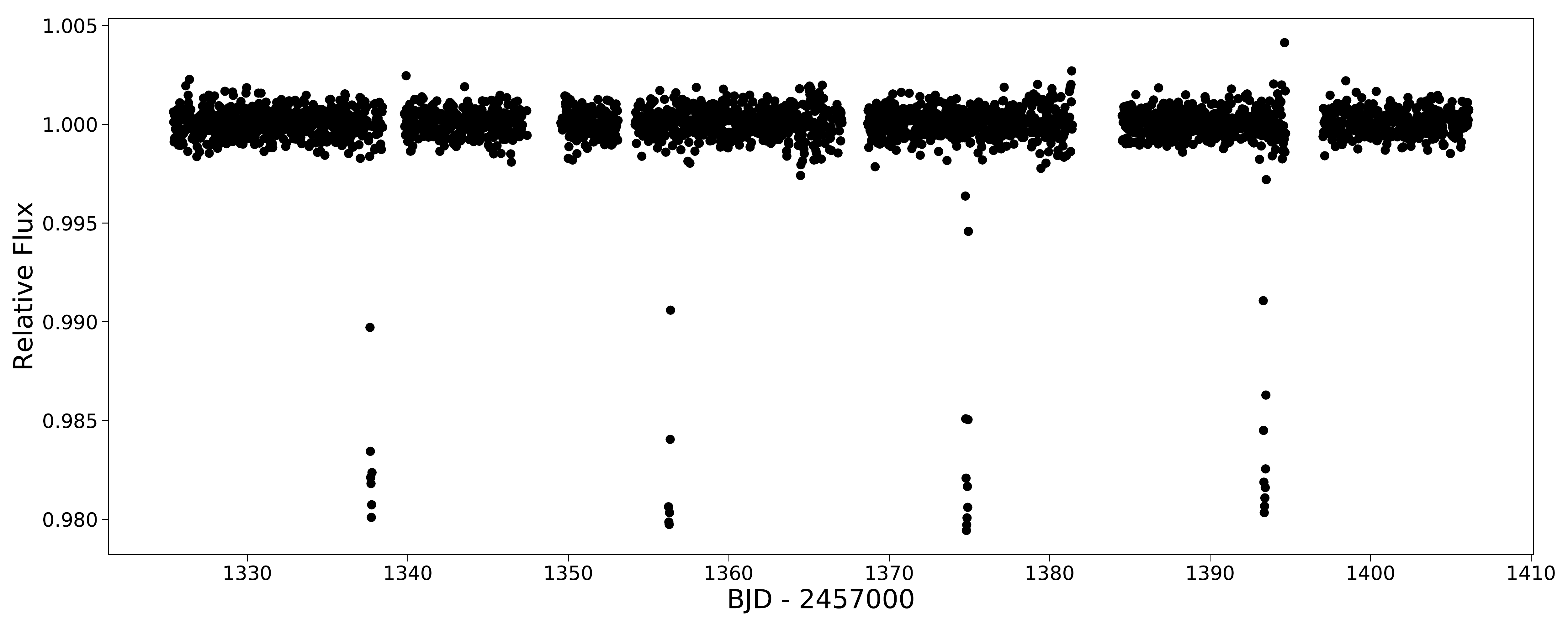}
\end{center}
\caption{30-minute cadence \tess\ light curve of \target\ with 4 clear transit events.}
\label{tess_lc_220}
\end{figure*}
\begin{figure*}
\includegraphics[width=\columnwidth]{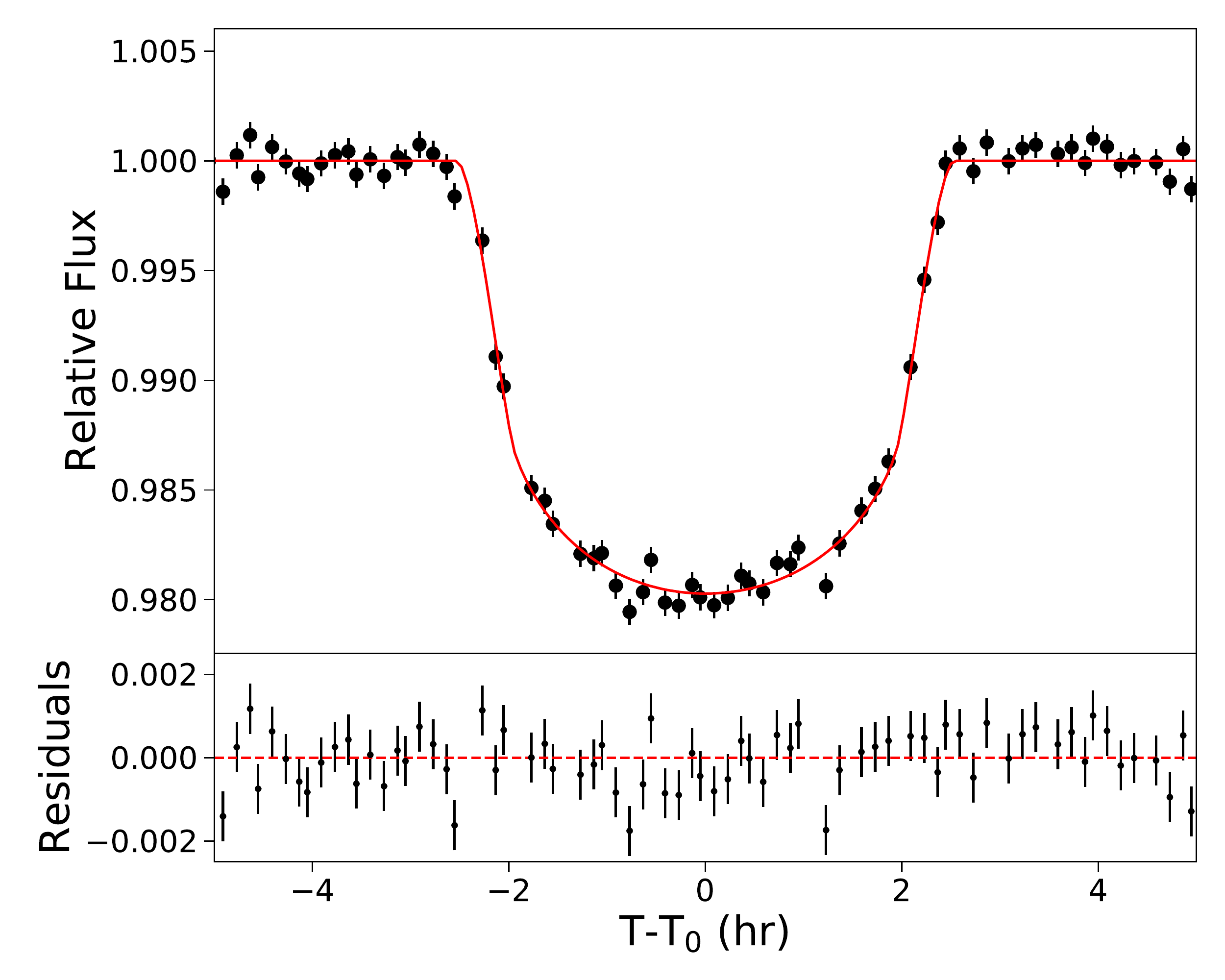}
\includegraphics[width=\columnwidth]{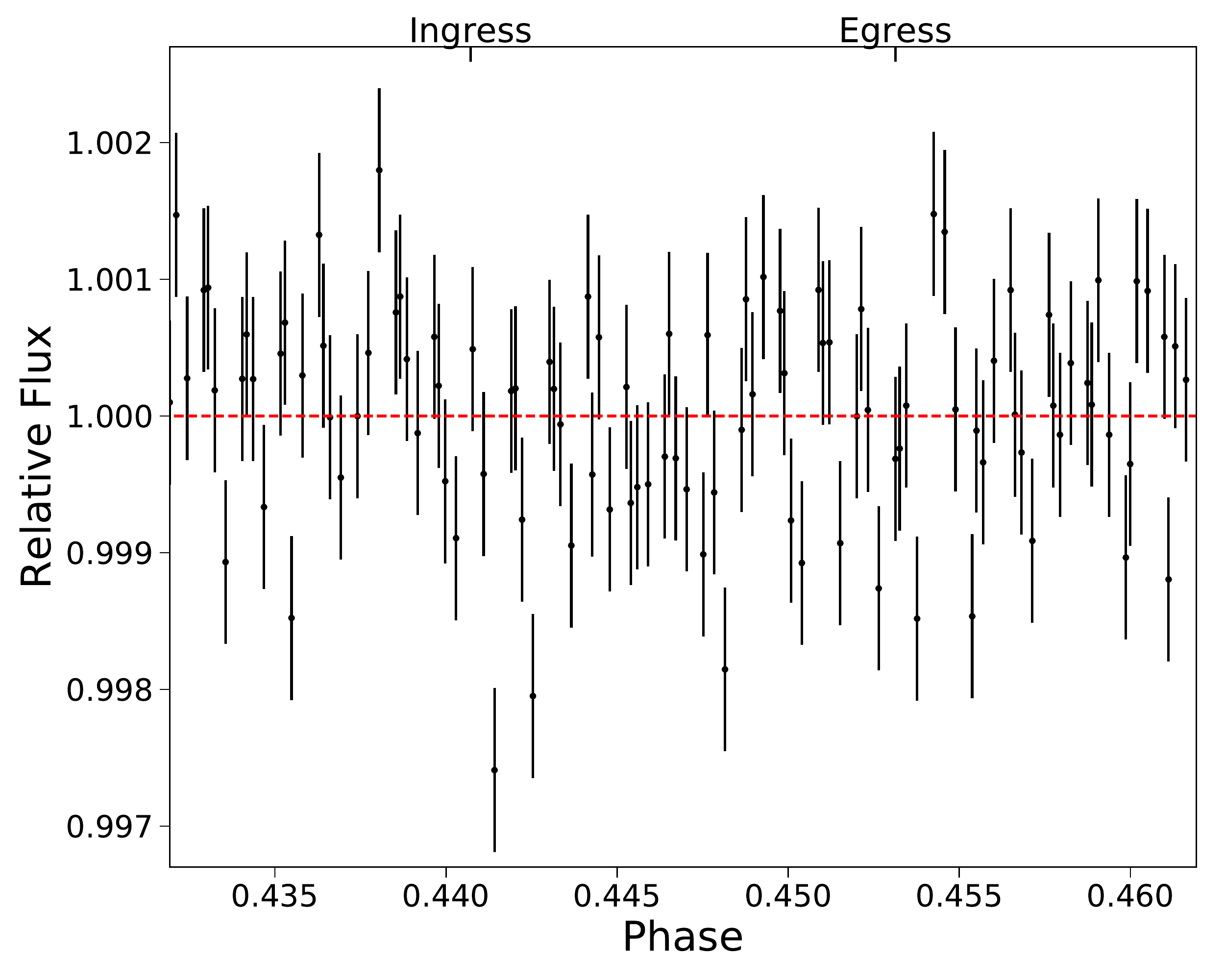}
\caption{Zoom-in of the phased \tess\ light curve of \target, centered on mid-transit (left) and the secondary eclipse (right). In the left panel, the red line is the fitted transit model, and the residual (data - model) are shown at the bottom. In the right panel, the dashed red line is the median relative flux value out of transit.}
\label{tess_lc_sec_220}
\end{figure*}

TIC 220568520 (\textit{V} = 12.039 mag) was observed by \tess\ Camera 3 during Sectors 1 through 3, from UT 2018 July 25 to UT 2018 October 18. 
The target's light curve was derived from the FFIs, with a 30 minute exposure time, by QLP where transits were identified with a high signal to noise ratio (SNR) every 18.2 days. The \tess\ light curves are shown in Figures \ref{tess_lc_220} and \ref{tess_lc_sec_220}. The decreased scatter in Figures \ref{tess_lc_220} and \ref{tess_lc_sec_220} relative to Figures \ref{tess_lc_694} and \ref{tess_lc_sec_694} is due to the longer integration time, which decreases the noise per exposure. 




\def\arraystretch{1.15}
\begin{deluxetable*}{lccc}{!h}
\tablewidth{0pc}
\tabletypesize{\scriptsize}
\tablecaption{
    Target Information
    \label{tab:info}
}
\tablehead{
    \multicolumn{1}{c}{Parameter} &
    \multicolumn{1}{c}{TOI 694} &
    \multicolumn{1}{c}{TIC 220568520} &
    \multicolumn{1}{c}{Source}
}
\startdata
TIC & 55383975 &  220568520 & TIC V8$^a$\\
R.A. & 05:09:32.06 & 03:05:08.58 & Gaia DR2$^b$ \\
Dec. & -64:01:33.9 &  -62:51:24.56 & Gaia DR2$^b$ \\
$\mu_{ra}$ (mas yr$^{-1}$)  & 6.468 $\pm$ 0.044 & 16.381 $\pm$ 0.038 & Gaia DR2$^b$ \\
$\mu_{dec}$ (mas yr$^{-1}$) & 25.987 $\pm$ 0.054 & 14.579 $\pm$ 0.036 & Gaia DR2$^b$ \\
Parallax (mas) & 4.438 $\pm$ 0.025 & 4.087 $\pm$ 0.021 & Gaia DR2$^b$ \\
Epoch & 2015.5 &  2015.5 & Gaia DR2$^b$\\
$B$ (mag)    & 12.761 $\pm$ 0.015 &  13.326 $\pm$ 0.045 & AAVSO DR9$^c$ \\
$V$ (mag)    & 11.963 $\pm$ 0.069 & 12.039 $\pm$ 0.05 & AAVSO DR9$^c$ \\
$Gaia$ (mag) & 11.7733 $\pm$ 0.00025 & 11.83746 $\pm$ 0.00023 & Gaia DR2$^b$ \\
$B_P$ (mag) & 12.2037 $\pm$ 0.0017 &  12.82419 $\pm$ 0.00075 & Gaia DR2$^b$ \\
$R_P$ (mag) & 11.21367 $\pm$ 0.00092 &  11.29765 $\pm$ 0.00055 & Gaia DR2$^b$ \\
\tess\ (mag) & 11.2595 $\pm$ 0.006 &  11.3458 $\pm$ 0.006 & TIC V8$^a$\\
$J$ (mag)    & 10.616 $\pm$ 0.024 &  10.687 $\pm$ 0.026 & 2MASS$^d$ \\
$H$ (mag)    & 10.207 $\pm$ 0.021 &  10.333 $\pm$ 0.022 & 2MASS$^d$ \\
$K_S$ (mag)  & 10.108 $\pm$ 0.019 &  10.270 $\pm$ 0.023 & 2MASS$^d$ 
\enddata
\tablenotetext{a}{\cite{stassun_tic_18}.}
\tablenotetext{b}{\cite{gaia18}.}
\tablenotetext{c}{\cite{henden16}.}
\tablenotetext{d}{\cite{cutri03}.}
\label{target_info}
\end{deluxetable*}

\subsection{Spectroscopic follow-up}
\label{sec:rv_data}

\begin{figure}[!h]
\includegraphics[width=\columnwidth]{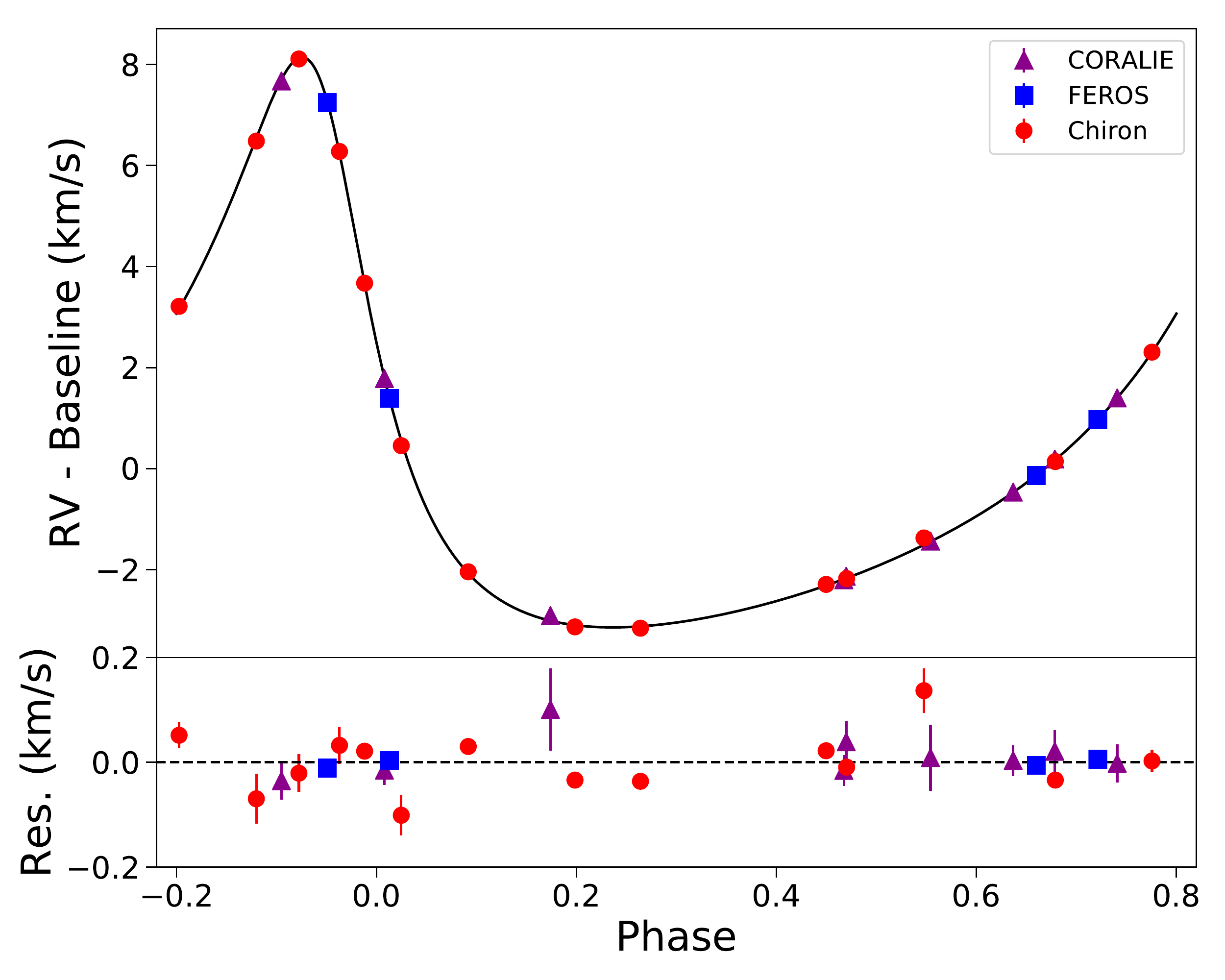}
\caption{Radial velocity curve of TOI 694 phase folded to the orbital period of the companion. The different colors and markers correspond to the different instruments used to obtain the data, and the black line denotes the best-fit model. The residuals for each instrument are shown in the bottom panel.}
\label{rv_data_fit_694}
\end{figure}

\begin{figure}[!h]
\includegraphics[width=\columnwidth]{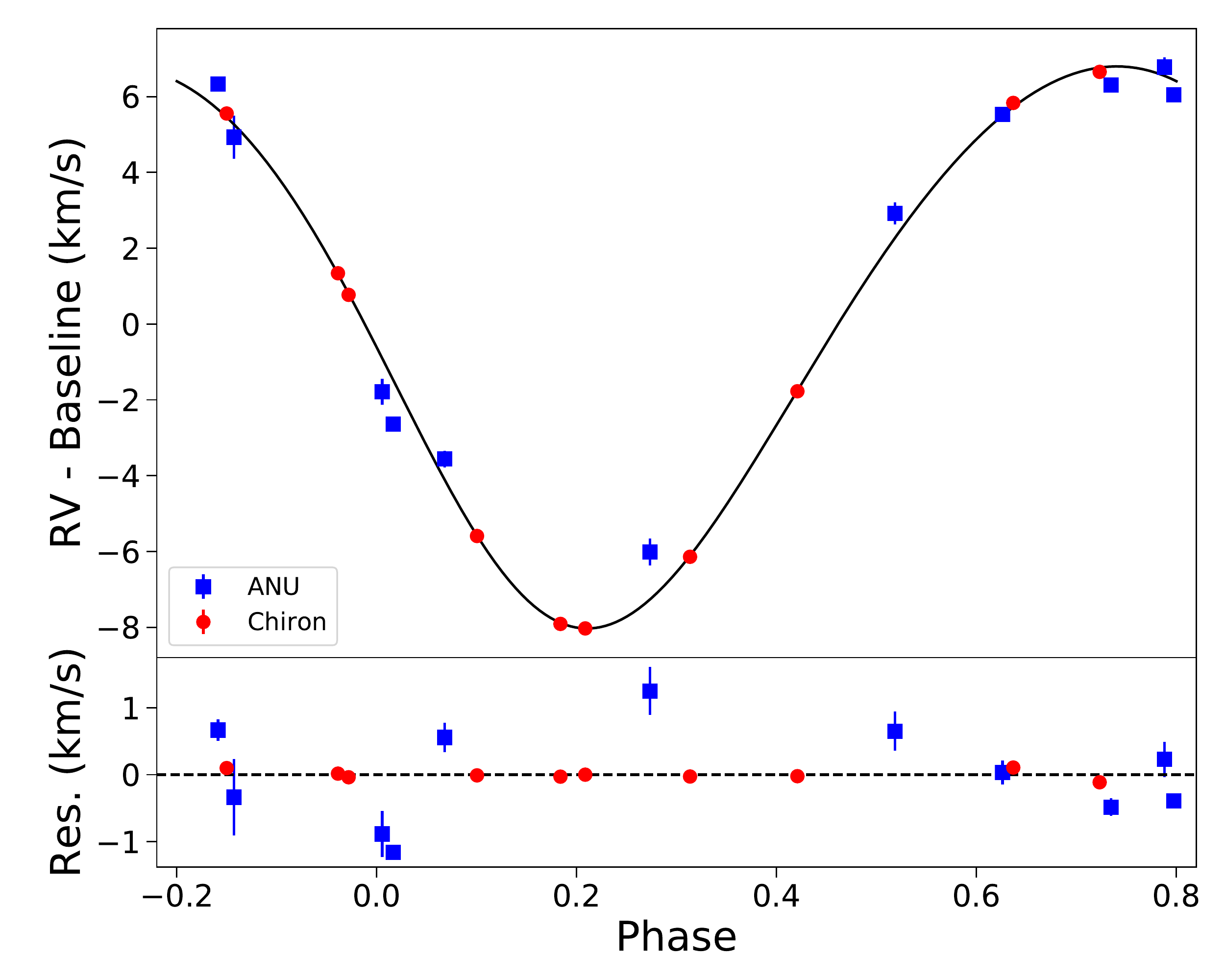}
\caption{Radial velocity curve of TIC 220568520 phase folded to the orbital period of the companion. The different colors and markers correspond to the different instruments used to obtain the data, and the black line denotes the best-fit model. The residuals for each instrument are shown in the bottom panel.}
\label{rv_data_fit_220}
\end{figure}

To characterize the stellar properties of the primary star and measure the mass of the transiting companion we obtained series of spectroscopic observations with the 1.5\,m SMARTS/CHIRON, ANU 2.3\,m/Echelle, Euler 1.2\,m/CORALIE, and MPG/ESO 2.2\,m/FEROS facilities. The radial velocity (RV) measurements are summarized in Table \ref{rv_data} and shown in Figures \ref{rv_data_fit_694} and \ref{rv_data_fit_220} for TOI~694 and \target, respectively.

For both objects we obtained time series radial velocity with the 1.5\,m SMARTS/CHIRON facility \citep{tokovinin2013_rv}, located at Cerro Tololo Inter-American Observatory (CTIO), Chile. The spectra were obtained with CHIRON in the fiber mode, with a spectral resolving power of $R\sim 25,000$ over the wavelength region of 4100\,\AA\, to 8700\,\AA. These observations were used to constrain the systems' RV orbit. Fourteen CHIRON spectra were obtained for TOI 694 from UT 2019 October 7 to UT 2020 February 17, and ten spectra were obtained for TIC 220568520 from 2019 August 31 to 2019 October 7. RVs were measured from each spectrum by modeling their rotational line profiles, derived via a least-square deconvolution \citep{donati1997} of the observed spectrum against a non-rotating synthetic template generated with the ATLAS-9 model atmospheres \citep{castelli2004}. The derived RVs are listed in Table~\ref{rv_data}. In addition, for each target a single CHIRON spectrum was observed in the slicer mode, with a resolving power of $R \sim 80,000$, to be used for spectral characterization of the primary star, described in Section \ref{stellarparams}.


For TIC 220568520 we measured RVs with the Australian National University (ANU) 2.3\,m/Echelle spectrograph at Siding Spring Observatory (New South Wales, Australia) prior to obtaining the CHIRON spectra. The ANU 2.3\,m/Echelle is a slit-fed medium resolution spectrograph on the ANU 2.3\,m telescope, located at Siding Spring Observatory, Australia. The spectrograph has a spectral resolving power of $\lambda / \Delta \lambda \equiv R \sim 23,000$ over the wavelength range of 3900\,\AA\, to 6700\,\AA. A total of eleven spectra were obtained for TIC 220568520 with the ANU 2.3\,m from UT 2019 January 17 to UT 2019 March 4th, at an average signal-to-noise ratio of $\sim 50$ per resolution element over the Mg b Triplet wavelength region for each observation. The spectra were reduced and extracted based on the procedure described in \citet{zhou2014_rv}, and are listed in Table \ref{rv_data}.

We obtained nine observations of TOI-694 with the high resolution spectrograph CORALIE on the Swiss 1.2 m Euler telescope at La Silla Observatory, Chile \citep{Queloz2001} between UT 2019 October 17 and UT 2020 January 7. CORALIE has a resolution of $R$ $\sim$ 60,000 and is fed by two fibers: a 2 arcsec on-sky science fiber encompassing the star and another fiber that can either connect to a Fabry-P\'erot etalon for simultaneous wavelength calibration or on-sky for background subtraction of sky flux. We observed TOI-694 in the simultaneous Fabry-P\'erot wavelength calibration mode using an exposure time of 1800 seconds. The spectra were reduced with the CORALIE standard reduction pipeline and RVs were computed for each epoch by cross-correlating with a binary G2 mask \citep{Pepe2002}. 

TOI-694 was observed also with the Fiber-fed, Extended Range, Echelle Spectrograph \citep[FEROS;][]{kaufer1999}, mounted on the MPG/ESO 2.2\,m telescope at La Silla, Chile. observations were done by the WINE collaboration, which is focused on the systematic characterization of warm giant planets with \tess\ \citep{brahm2019,jordan2020}. Four spectra of TOI-694 were obtained between UT 2019 February 28 and UT 2019 March 17 with an exposure time of 900 sec and a SNR ranging from 53 to 76. Observations were performed with the simultaneous calibration technique, where a comparison fiber is used to trace the instrumental variations during the science exposure by registering the spectrum of a ThAr lamp. FEROS data were reduced and processed with the \texttt{CERES} pipeline \citep{ceres} which uses the optimal extraction routines presented by \citet{marsh1989} and delivers precise RVs and bisector span measurements, which are presented in Table \ref{rv_data}.


\begin{deluxetable}{lccc}[!h]
\tablewidth{0pc}
\tabletypesize{\scriptsize}
\tablecaption{
    Radial Velocities
    \label{tab:rvs}
}
\tablehead{
    \multicolumn{1}{c}{Time} &
    \multicolumn{1}{c}{RV\tablenotemark{a}}    &
    \multicolumn{1}{c}{Error} &
    \multicolumn{1}{c}{Instrument} \\
    \multicolumn{1}{c}{BJD} &
    \multicolumn{1}{c}{\kms}    &
    \multicolumn{1}{c}{\kms} &
    \multicolumn{1}{c}{}
    }
\startdata
\multicolumn{4}{c}{TOI 694\tablenotemark{b}} \\
\hline

2458542.63751 & 21.4803 & 0.0076 & FEROS \\
2458545.58904 & 22.5943 & 0.0071 & FEROS \\
2458556.62596 & 28.8646 & 0.0095 & FEROS \\
2458559.60797 & 23.0099 & 0.0077 & FEROS \\
2458763.86992 & 16.2069 & 0.0149 & CHIRON \\
2458772.79219 & 17.0740 & 0.0104 & CHIRON  \\
2458773.77718 & 17.1886 & 0.0145 & CHIRON  \\
2458783.80427 & 19.5069 & 0.0068 & CHIRON  \\
2458789.75347 & 22.5803 & 0.0252 & CHIRON  \\
2458798.66760 & 23.0389 & 0.0159 & CHIRON  \\
2458803.64710 & 17.3233 & 0.0147 & CHIRON  \\
2458808.77574 & 16.2327 & 0.0156 & CHIRON  \\
2458873.59450 & 17.9937 & 0.0422 & CHIRON  \\
2458884.55191 & 21.6734 & 0.0214 & CHIRON  \\
2458889.56772 & 25.8525 & 0.0477 & CHIRON  \\
2458891.61208 & 27.4781 & 0.0359 & CHIRON  \\
2458893.56327 & 25.6463 & 0.0351 & CHIRON  \\
2458896.53063 & 19.8216 & 0.0386 & CHIRON  \\
2458773.75679 & 19.5107 & 0.0396 & CORALIE \\
2458777.80659 & 20.2075 & 0.0631 & CORALIE \\
2458781.77898 & 21.1769 & 0.0293 & CORALIE \\
2458783.78105 & 21.8304 & 0.0406 & CORALIE \\
2458786.77769 & 23.0417 & 0.0366 & CORALIE \\
2458821.69504 & 19.4387 & 0.0295 & CORALIE \\
2458842.72232 & 29.3174 & 0.0362 & CORALIE \\
2458847.67201 & 23.4254 & 0.0283 & CORALIE \\
2458855.64763 & 18.7321 & 0.0783 & CORALIE \\
\\
\multicolumn{4}{c}{TIC 220568520\tablenotemark{c}} \\
\hline
2458500.97894 & 30.5575 & 0.0318 & ANU 2.3 m/Echelle \\
2458502.09007 & 29.4386 & 0.5716 & ANU 2.3 m/Echelle \\
2458505.04661 & 21.8699 & 0.0795 & ANU 2.3 m/Echelle \\
2458506.00173 & 20.9493 & 0.2224 & ANU 2.3 m/Echelle \\
2458532.91371 & 27.4253 & 0.2931 & ANU 2.3 m/Echelle \\
2458534.91425 & 30.0442 & 0.1773 & ANU 2.3 m/Echelle \\
2458536.92492 & 30.8171 & 0.1309 & ANU 2.3 m/Echelle \\
2458537.91918 & 31.2864 & 0.2634 & ANU 2.3 m/Echelle \\
2458538.90685 & 30.8376 & 0.1639 & ANU 2.3 m/Echelle \\
2458541.95366 & 22.7282 & 0.3439 & ANU 2.3 m/Echelle \\
2458546.92590 & 18.4931 & 0.3579 & ANU 2.3 m/Echelle \\
2458726.90477 & 24.0542 & 0.0239 & CHIRON \\
2458730.83770 & 15.3734 & 0.0217 & CHIRON \\
2458740.84783 & 29.9363 & 0.0211 & CHIRON \\
2458747.84631 & 17.6927 & 0.0115 & CHIRON \\
2458749.85537 & 15.2544 & 0.0140 & CHIRON \\
2458751.80097 & 17.1443 & 0.0172 & CHIRON \\
2458753.79406 & 21.5104 & 0.0119 & CHIRON \\
2458757.80128 & 29.1168	& 0.0187 & CHIRON \\
2458761.75638 & 28.8378	& 0.0263 & CHIRON \\
2458763.82222 & 24.6238 & 0.0131 & CHIRON
\enddata
\tablenotetext{a}{All radial velocities are barycentric.}
\tablenotetext{b}{The Gaia DR2 RV is 20.70 $\pm$ 2.22 \kms \citep{gaia2018}.}
\tablenotetext{c}{The Gaia DR2 RV is 28.09 $\pm$ 1.79 \kms \citep{gaia2018}.}
\label{rv_data}
\end{deluxetable}

\subsection{High Angular Resolution Imaging with SOAR HRCam}
\label{sec:imaging}

We used high angular resolution imaging in order to look for stars close to the targets' position, within an angular separation of 1--2\arcsec, which cannot be identified with regular seeing-limited imaging. If they exist, the small separation might prevent them from being detected by Gaia and included in the TIC \citep{stassun_tic_18}, leading to an inaccurate estimate of the extent by which light from the target is blended with light from nearby stars, leading in turn to a biased estimate of the intrinsic transit depth. It is also possible that nearby stars are the source of the variability seen in \tess\ data.

\begin{figure}[!h]
\includegraphics[width=\columnwidth]{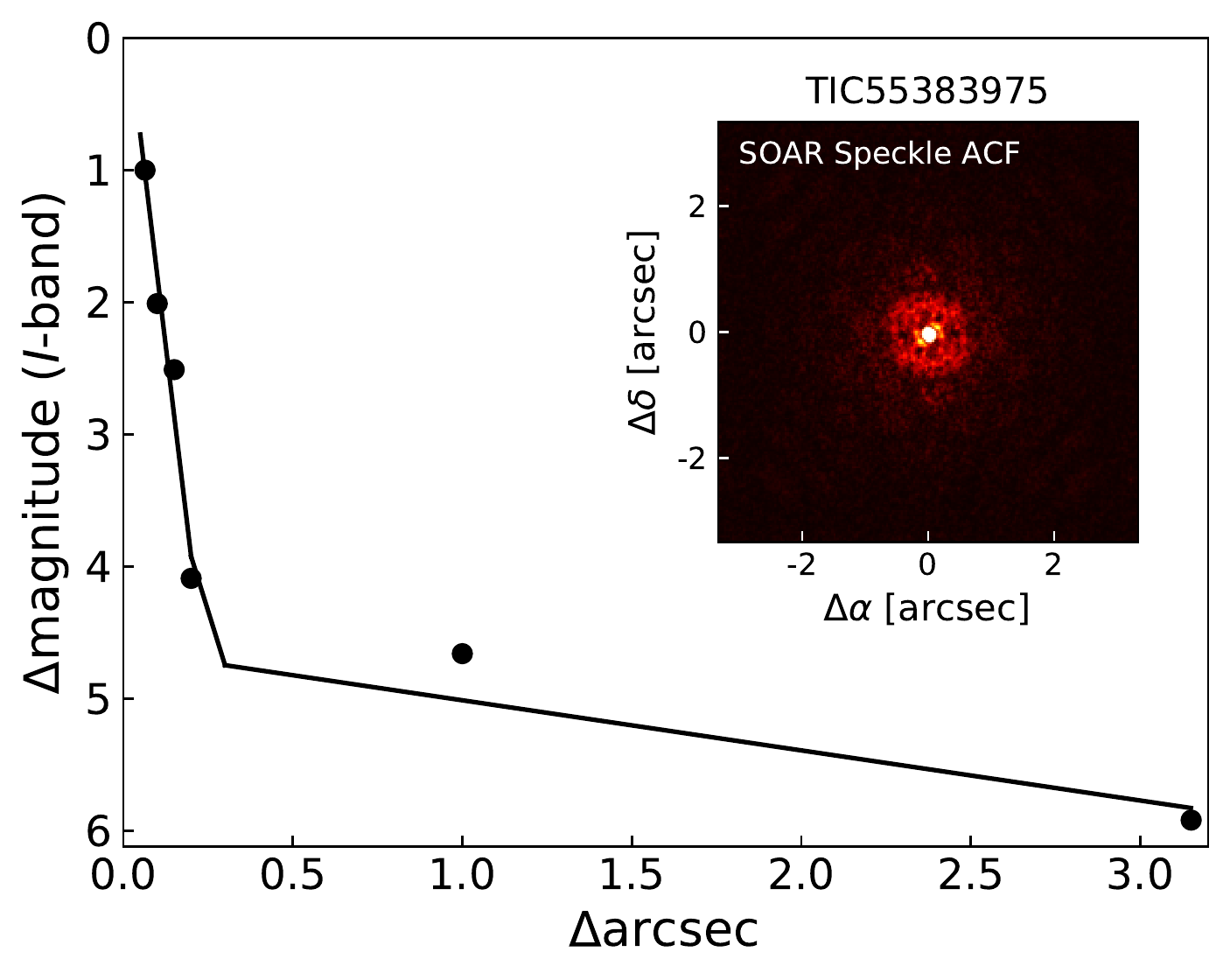}
\caption{Contrast curves showing the 5-$\sigma$ detection sensitivity and speckle auto-correlation functions obtained in I-band using SOAR/HRCam for TOI 694 (TIC 55383975).}
\label{speckle_fig_694}
\end{figure}

\begin{figure}[!h]
\includegraphics[width=\columnwidth]{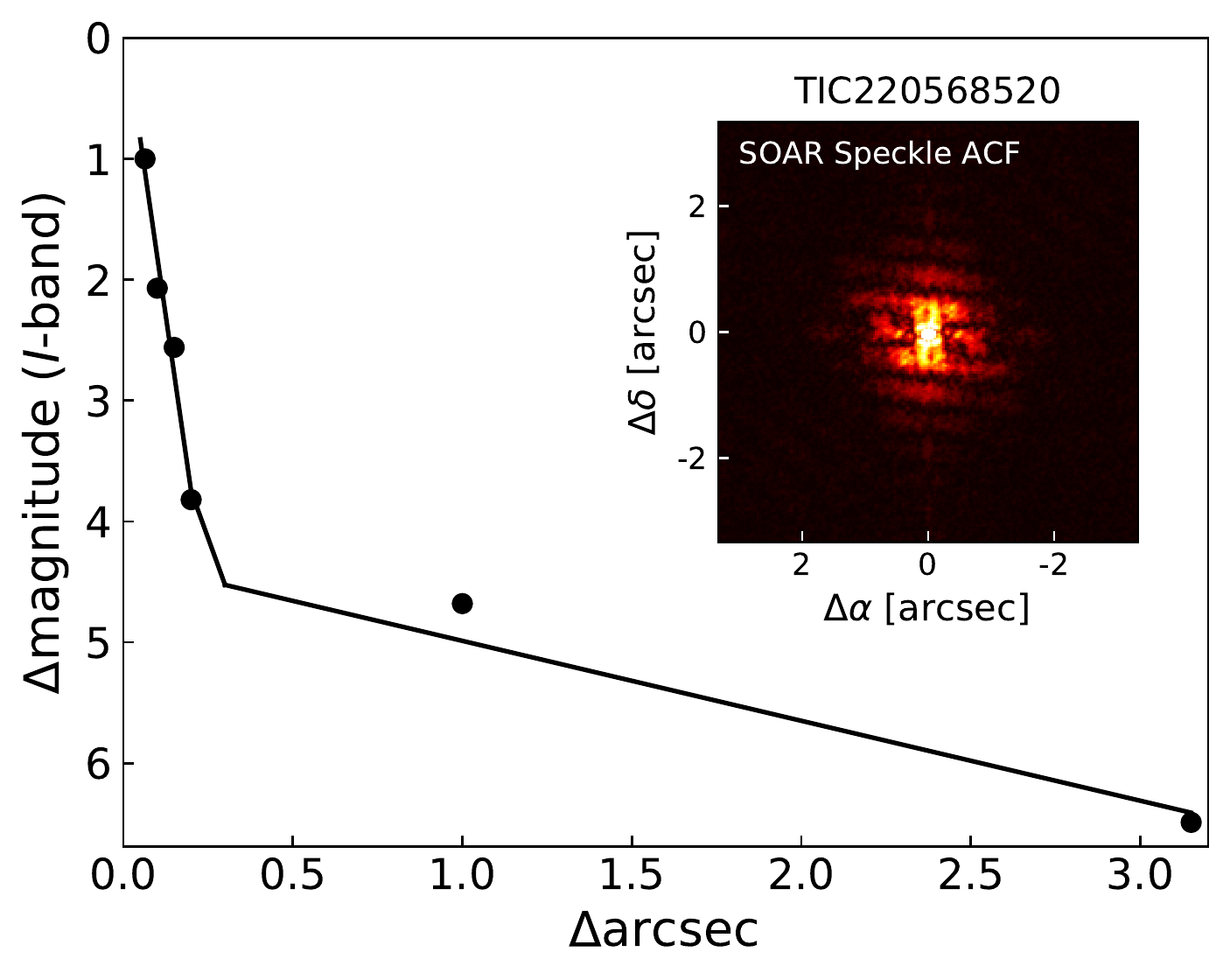}
\caption{Contrast curves showing the 5-$\sigma$ detection sensitivity and speckle auto-correlation functions obtained in I-band using SOAR/HRCam for TIC 220568520.}
\label{speckle_fig_220}
\end{figure}

We observed both targets with the high resolution camera \citep[HRCam,][]{tokovinin_soar2018} mounted on the southern astrophysical research (SOAR) 4.1\,m telescope, in Cerro Pach\'{o}n, Chile. HRCam uses the speckle interferometry technique in a visible bandpass similar to that of \tess. A detailed description of HRCam observations of \tess\ targets is available in \citet{ziegler_soar2020}. 
We observed TOI 694 on UT 2019 July 14 and TIC 220568520 on UT 2019 November 9. For both targets we detected no nearby sources down to $\Delta$mag $\approx$ 4.5 mag at 0.3\arcsec\ and $\Delta$mag $\approx$ 5.0 at 1.0\arcsec. The 5-$\sigma$ SOAR/HRCam detection sensitivity and the speckle auto-correlation function are plotted in Figures \ref{speckle_fig_694} and \ref{speckle_fig_220}.

\section{Data Analysis}
\label{sec:data_analysis}

\subsection{Primary Star Parameters}
\label{stellarparams}

\subsubsection{MESA Isochrones and Stellar Tracks Analysis}

For the primary star in each system we derived initial values of the spectral parameters, \teff, \logg, and \vsini, by matching each CHIRON spectrum against a library of $\sim$10,000 observed spectra classified by the Stellar Parameter Classification routine \citep{buchhave2012}.

We then used the spectroscopic parameters along with the Gaia DR2 parallax and magnitudes ($G,\, B_P,\, R_P$), 2MASS magnitudes ($J, H, K_S$), and AAVSO magnitudes ($B, V$) to perform an isochrone fit in order to further constrain the spectroscopic parameters and derive physical parameters for the primary stars. The spectroscopic parameters, parallax, and magnitudes are used as priors to determine the goodness of fit. We use the \emph{isochrone} package \citep{isochrone_python} to generate the isochrone models used to sample the stellar parameters and find the best-fit parameters by using an MCMC routine. The routine consists of 40 independent walkers each taking 25000 steps, of which the first 2000 are discarded as burn-in.

The fitted spectroscopic parameters and derived physical parameters for the primary stars in each of the two systems studied here are reported in Table \ref{fitted_derived_parameters}. Both primary stars are similar to each other, and similar to the Sun (within 1--2 $\sigma$) albeit with higher metallicities.

\subsubsection{Spectral Energy Distribution Analysis}



\begin{figure}[!ht]
    \centering
    \includegraphics[width=\linewidth,trim=100 125 90 90,clip]{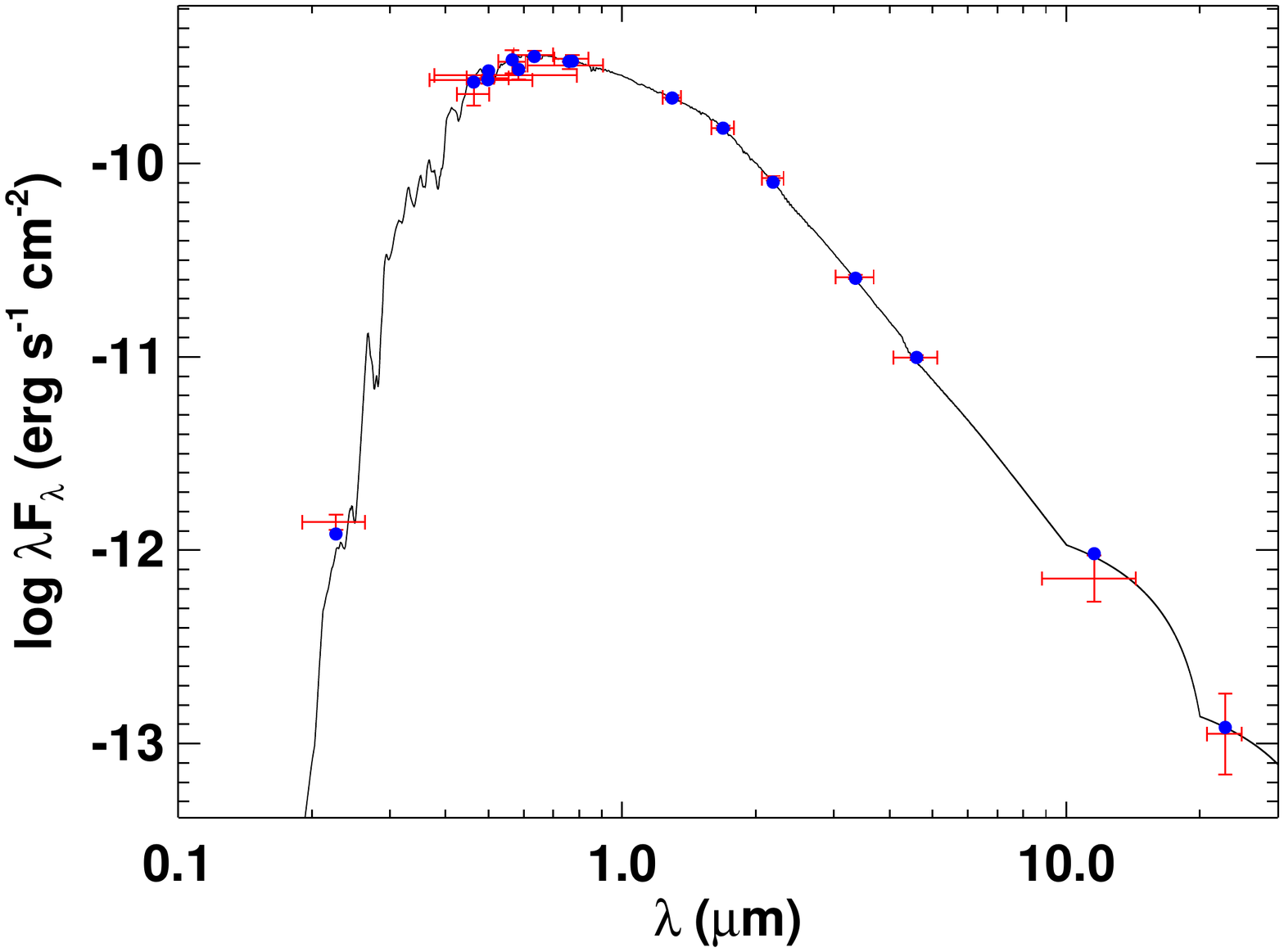}
    \includegraphics[width=\linewidth,trim=100 75 90 90,clip]{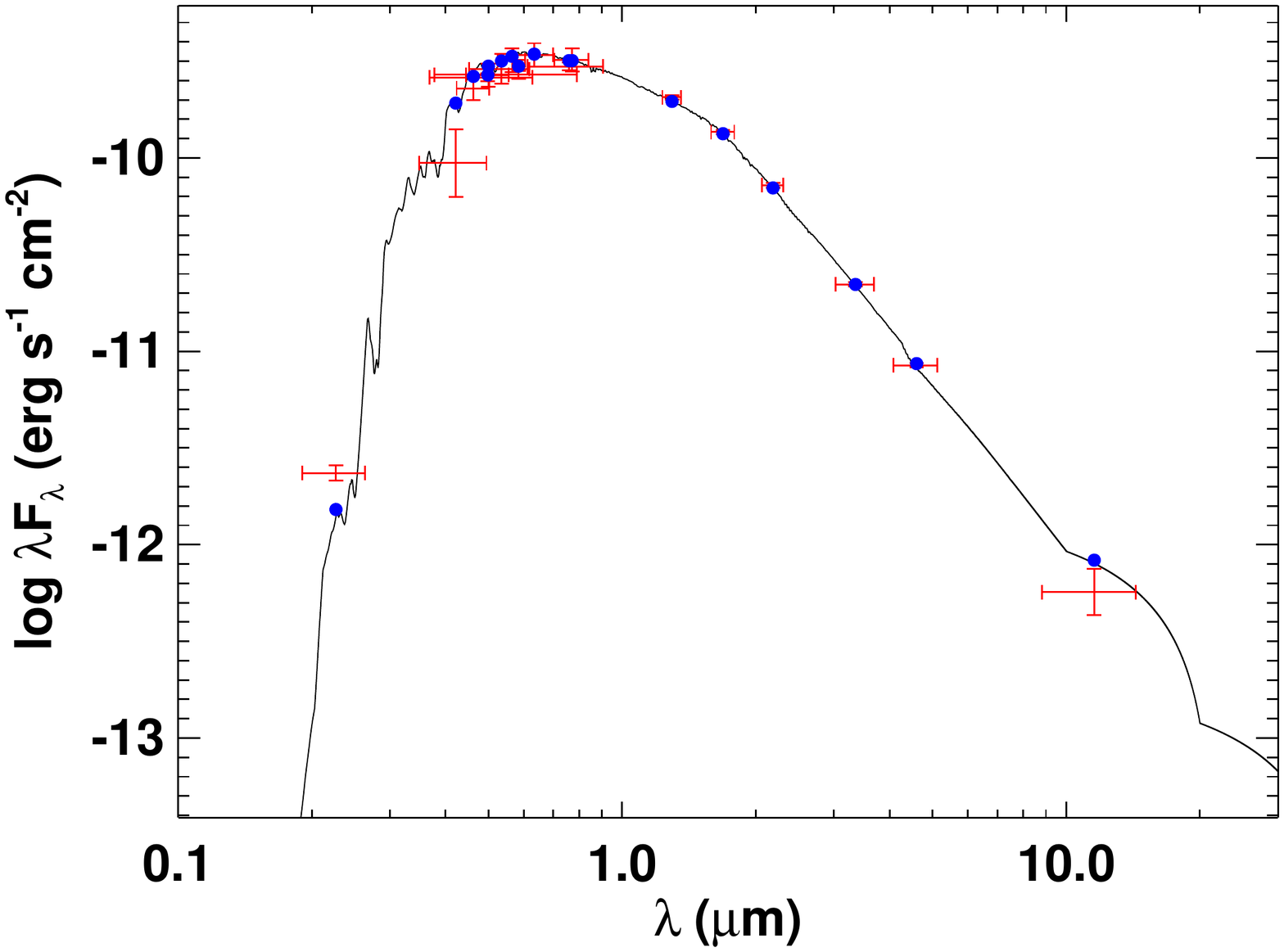}
    \caption{Spectral energy distribution (SED) for TOI~694 (top) and TIC~220568520 (bottom). Red symbols represent the observed photometric measurements, where the horizontal bars represent the effective width of the passband. Blue symbols are the model fluxes from the best-fit Kurucz atmosphere model (black). 
\label{fig:sed}}
\end{figure}

For both TOI~694 and TIC~220568520 
we performed an analysis of the broadband spectral energy distribution (SED) together with the {\it Gaia\/} DR2 parallax in order to determine an empirical measurement of the stellar radius, following the procedures described in \citet{Stassun:2016,Stassun:2017,Stassun:2018}. We pulled the $BVgri$ magnitudes from APASS, the $JHK_S$ magnitudes from {\it 2MASS}, the W1--W4 magnitudes from {\it WISE}, and the $G G_{\rm BP} G_{\rm RP}$ magnitudes from {\it Gaia}. We also considered the {\it GALEX\/} NUV flux for evidence of chromospheric activity. Together, the available photometry spans the full stellar SED over the wavelength range of 0.2--22~$\mu$m (see Figure~\ref{fig:sed}). 

We performed a fit using Kurucz stellar atmosphere models, with the priors on effective temperature (\teff), surface gravity (\logg), and metallicity (\feh) from the analysis of the CHIRON spectra. The remaining free parameter is the extinction ($A_V$), which we limited to the maximum permitted for the star's line of sight from the \citet{Schlegel:1998} dust maps. The resulting fits, plotted in Figure~\ref{fig:sed}, have a reduced $\chi^2$ of 1.3 and 2.1, and $A_V = 0.12\pm0.02$ and $0.06\pm0.02$, for TOI~694 and TIC~220568520, respectively. Integrating the model SED gives the bolometric flux at Earth of $F_{\rm bol} = 5.199 \pm 0.060 \times 10^{-10}$ erg~s$^{-1}$~cm$^{-2}$ and $4.77 \pm 0.11 \times 10^{-10}$ erg~s$^{-1}$~cm$^{-2}$, respectively. Taking the $F_{\rm bol}$ and \teff\ together with the {\it Gaia\/} parallax, adjusted by $+0.08$~mas to account for the systematic offset reported by \citet{StassunTorres:2018}, gives the stellar radii as \rstar $= 0.992 \pm 0.037$~\rsun\, and $0.992 \pm 0.038$~\rsun, respectively. 

\subsubsection{Stellar Mass via Radius and Surface Gravity}
\label{sec:hostmass}
The empirical stellar radii determined above affords an opportunity to estimate the stellar masses empirically as well, via the spectroscopically determined surface gravity (\logg $= 4.45\pm0.10$ and $4.34\pm0.10$, respectively). For TOI~694 and TIC~220568520 we obtain \mstar $= 1.01 \pm 0.13$~\msun\ and $0.79 \pm 0.19$~\msun, respectively. These are similar to the values estimated via the eclipsing-binary based relations of \citet{Torres:2010}, which give \mstar $=1.00 \pm 0.06$~\msun\ and $1.09 \pm 0.07$~\msun, respectively. For TOI~694, these values are nearly identical. For TIC~220568520, they differ by less than 2 $\sigma$.

\subsubsection{Stellar Age via Gyrochronology}
\label{sec:hostrot}
We estimate each star's age from its rotational period, $P_{\rm rot}$, which we calculate from the spectroscopic \vsini\ together with the empirically determined radius above, assuming $I = 90^{\circ}$. For TOI~694 and TIC~220568520 we obtain $P_{\rm rot} \approx 24.1 \pm 5.6$~d and $15.5 \pm 2.3$~d, respectively.
From the rotation-activity-age relations of \citet{MamajekHillenbrand:2008}, we obtain from the $P_{\rm rot}$ and the stellar $B-V$ colors ages of $\tau = 3.8 \pm 0.4$~Gyr and $2.3 \pm 0.3$~Gyr, respectively.

\subsection{Simultaneous Transit and RV Fit}

In order to derive the orbital parameters and companion's radius and mass, we performed model fitting using {\scshape allesfitter}\footnote{\url{https://github.com/MNGuenther/allesfitter}} \citep{allesfitter2019, gunther2020}, enabling a joint analysis of the \tess\ transit light curve and the RV orbit. {\scshape allesfitter} is publicly available, and provides an environment to analyze light curves and RVs of binary star and star-planet systems. It is based on various public packages including {\scshape ellc} for light curve and RV modeling (\citealt{ellc_maxted}) and {\scshape emcee} for Markov Chain Monte Carlo sampling (\citealt{emcee2013}).


We fit the following parameters:
\begin{itemize}
    \itemsep0pt 
    \item quadratic stellar limb-darkening parameters $q_1$ and $q_2$, using the transformation from \citet{Kipping_2013}, with Gaussian priors centered on values derived from \citet{claret2017}. We chose the Gaussian prior width (1 $\sigma$) to be 0.1, which reflects the uncertainties on the host stars parameters.
    \item radius ratio, $R_2/R_1$, where 1 denotes the primary star and 2 the secondary, with uniform prior from 0 to 1,
    \item sum of radii divided by the orbital semi-major axis, $(R_1 + R_2) / a$, with uniform prior from 0 to 1,
    \item cosine of the orbital inclination, $ \cos{i}$, with uniform prior from 0 to 1, 
    \item orbital period, $ P $ with uniform prior from 0 to $10^{12}$ days,
    \item Primary eclipse epoch, $ T_{0} $, with uniform prior from 0 to $10^{12}$ days,
    \item RV semi-amplitude, $K$, with uniform prior from 0 to 50 \kms,
    \item eccentricity parameters $\sqrt{e}\, cos\, \omega$ and $\sqrt{e}\, sin\, \omega$, each with uniform prior from -1 to 1, where $e$ is the orbital eccentricity and $\omega$ the argument of periastron,
    \item RV zero point, $\gamma_c$, for each of the RV data sets,
    \item RV jitter terms, $\sigma_c$, for each of the RV data sets,
    \item white noise scaling term for the \tess\ data, $\sigma_{c, TESS}$.

\end{itemize}

Initial guesses for the values of $q_1$ and $q_2$ were obtained by matching the spectroscopic parameters of each primary star to the closest values of the coefficients $u_1$ and $u_2$ of the quadratic limb darkening law listed in \citet{claret2017}, and transforming them to the corresponding values of $q_1$ and $q_2$. Initial guesses for $R_2/R_1$, $(R_1 + R_2) / a$,  $\cos{i}$, $P$, and $T_{0}$ were obtained using values provided by the SPOC Data Validation Report for TOI 694 \citep{li2019_dvr} and QLP Data Validation Report for \target. Initial guesses for $K$ and the 4 $\gamma_c$ terms were obtained by visually inspecting each data set. We used an MCMC algorithm to explore the parameter space and determine the best-fit parameters. We initialized the MCMC with 100 walkers, performing 2 preliminary runs of 1000 steps per walker to obtain higher-likelihood initial guesses for the nominal run of 15000 steps per walker. We then discarded the first 2000 steps for each chain as burn-in phase before thinning the chains by a factor of 10 and calculating the final posterior distributions. The values and errors of the fitted and derived parameters listed in Table \ref{fitted_derived_parameters} are defined as the median values and 68\% confidence intervals of the posterior distributions, respectively. The best-fit transit model light curve and radial velocity model curve for TOI~694 are shown in Figures \ref{tess_lc_sec_694} and \ref{rv_data_fit_694}, respectively. For \target, they are shown in Figures \ref{tess_lc_sec_220} and \ref{rv_data_fit_220}.


\def\arraystretch{1.35}
\begin{deluxetable*}{lcccc}[!h]
\tablewidth{0pc}
\tabletypesize{\scriptsize}
\tablecaption{
    Fitted and derived Parameters
    \label{tab:params}
}
\tablehead{
    \multicolumn{1}{c}{~~~~~~~~~~Parameter~~~~~~~~~~} &
    \multicolumn{2}{c}{TOI 694} &
    \multicolumn{2}{c}{TIC 220568520} \\
                              &
    \multicolumn{1}{c}{Value} &
    \multicolumn{1}{c}{Error} &  
    \multicolumn{1}{c}{Value} &
    \multicolumn{1}{c}{Error}
}
\startdata
\multicolumn{3}{l}{\textit{Host star Parameters}} \\ 
\mpri\ (\msun)  & 0.967 & $^{+0.047}_{-0.040}$ & 1.030 & $^{+0.043}_{-0.042}$ \\
\rpri\ (\rsun)  & 0.998 & $^{+0.010}_{-0.012}$ & 1.007 & $^{+0.010}_{-0.009}$ \\ 
$\rho_1$\tablenotemark{a} (\gcmc) & 1.51 & $_{-0.13}^{+0.15}$ & 1.445 & $_{-0.074}^{+0.064}$ \\
\lpri\ (\lsun)  & 0.819 & $^{+0.047}_{-0.044}$ &  0.890 & $^{+0.047}_{-0.044}$ \\ 
\teff\ (K)      & 5496 & $^{+87}_{-81}$ & 5589 & $81$ \\ 
\feh\           & 0.21 & 0.08 & 0.26 & 0.07 \\
\logg           & 4.425 & $^{+0.028}_{-0.025}$ & 4.445 & $^{+0.023}_{-0.025}$ \\
Age (Gyr)       & 7.33 & $^{+2.92}_{-3.02}$ & 4.09 & $^{+2.60}_{-2.28}$ \\ 
$\vsini$ (\kms) & 2.18 & 0.5 & 3.4 & 0.5 \\ \\
\multicolumn{3}{l}{\textit{Fitted parameters}} \\
$q_{1, \tess}$ & 0.345 & $_{-0.070}^{+0.078}$ & 0.418 & $_{-0.074}^{+0.080}$ \\ 
$q_{2, \tess}$ & 0.308 & $_{-0.079}^{+0.084}$ & 0.374 & $0.071$  \\
$R_2 / R_1$    & 0.1145 & $_{-0.0013}^{+0.0012}$ & 0.1274 & $_{-0.0013}^{+0.0014}$ \\ 
$(R_1 + R_2) / a$ & 0.01960 & 0.00064 & 0.03791  & $_{-0.00056}^{+0.00070}$ \\ 
$\cos{i}$ & 0.0152 & $_{-0.0023}^{+0.0020}$ & 0.0075 & $_{-0.0040}^{+0.0030}$ \\ 
$T_0$ (BJD - 2,457,000) & 1366.77176 & 0.00064 & 1337.72060 & 0.00082 \\ 
$P$   $\mathrm{(d)}$ & 48.05131 & 0.00019 & 18.55769  &  0.00039   \\
$K$ (\kms) & 5.645 & 0.011 & 7.428 & $0.035$ \\
$\sqrt{e}\, cos\, \omega$ & 0.61438 & $_{-0.00078}^{+0.00072}$ & -0.2709 & 0.0080 \\
$\sqrt{e}\, sin\, \omega$ & 0.3768 & $_{-0.0015}^{+0.0016}$ & 0.151 & 0.016 \\ 
$\gamma_{c,ANU}$ (\kms) & - & - & 24.50 & 0.22 \\
$\gamma_{c,CHIRON}$ (\kms) & 19.368 & 0.013 & 23.285 & 0.028 \\
$\gamma_{c,FEROS}$ (\kms) & 21.6187 & 0.0062 & - & - \\
$\gamma_{c,CORALIE}$ (\kms) & 21.645 & 0.012 & - & - \\
ln $\sigma_{c,ANU}$ [\kms] & - & - & -0.37 & $_{-0.23}^{+0.21}$ \\
ln $\sigma_{c,CHIRON}$ [\kms] & -3.21 & 0.34 & -2.56 & $_{-0.30}^{+0.36}$ \\
ln $\sigma_{c,FEROS}$ [\kms] & -12.7 & 6.9 & - & - \\
ln $\sigma_{c,CORALIE}$ [\kms] & -13.5 & 6.5 & - & - \\
ln $\sigma_{c,TESS_{SC}}$\tablenotemark{b} & -6.242 & 0.012 & - & - \\
ln $\sigma_{c,TESS_{LC}}$\tablenotemark{c} & -7.21 & $_{-0.17}^{+0.20}$ & -7.313 & 0.055 \\ \\
\multicolumn{3}{l}{\textit{Derived Parameters}} \\ 
$i$ ($^{\circ}$) & 89.13 & $_{-0.11}^{+0.13}$ & 89.57 & $_{-0.17}^{+0.23}$ \\
$a$ (AU) & 0.2638 & $_{-0.0086}^{+0.0092}$ & 0.1391 & $_{-0.0027}^{+0.0025}$ \\
$b$\tablenotemark{d} & 0.497 & $_{-0.060}^{+0.048}$ & 0.211 & $_{-0.11}^{+0.081}$ \\ 
$T_\mathrm{tot}$\tablenotemark{e} (h) & 4.328 & $_{-0.035}^{+0.037}$ & 5.018 & 0.035 \\ 
$T_\mathrm{full}$\tablenotemark{f}  (h) & 3.179 & 0.068 & 3.834 & $_{-0.058}^{+0.050}$ \\ 
$e$ & 0.51946 & 0.00081 & 0.0964 & $_{-0.0030}^{+0.0034}$ \\
$\omega$ (deg) & 31.52 & $_{-0.12}^{+0.13}$ & 150.9 & 3.2 \\
\rsec\, (\rjup) & 1.111 & 0.017 & 1.248 & 0.018 \\
\msec\, (\mjup) & 89.0 & 5.3 & 107.2 & 5.2
\enddata
\tablenotetext{a}{Derived from simultaneous fit of \tess\, photometry and RV measurements.}
\tablenotetext{b}{\tess\ 2 minute cadence data.}
\tablenotetext{c}{\tess\ 30 minute cadence data.}
\tablenotetext{d}{Impact parameter.}
\tablenotetext{e}{From 1st to last (4th) contacts.}
\tablenotetext{f}{From 2nd to 3rd contacts.}
\label{fitted_derived_parameters}
\end{deluxetable*}

\section{Discussion}
\label{sec:dis}

From the simultaneous fits of the \tess\ transit photometry and RV data, we derive TOI 694 b mass and radius to be 89.0 $\pm$ 5.3 \mjup\ (0.0849 $\pm$ 0.0051 \msun) and 1.111 $\pm$ 0.017 \rjup\ (0.1142 $\pm$ 0.0017 \rsun), respectively, and TIC 220568520 b mass and radius to be 107.2 $\pm$ 5.2 \mjup\ (0.1023 $\pm$ 0.0050 \msun) and 1.248 $\pm$ 0.018 \rjup\ (0.1282 $\pm$ 0.0019 \rsun), respectively. We note that mass uncertainties are dominated by the primary star mass uncertainties and not the orbital parameters uncertainties.

\begin{figure*}
\includegraphics[width=17cm]{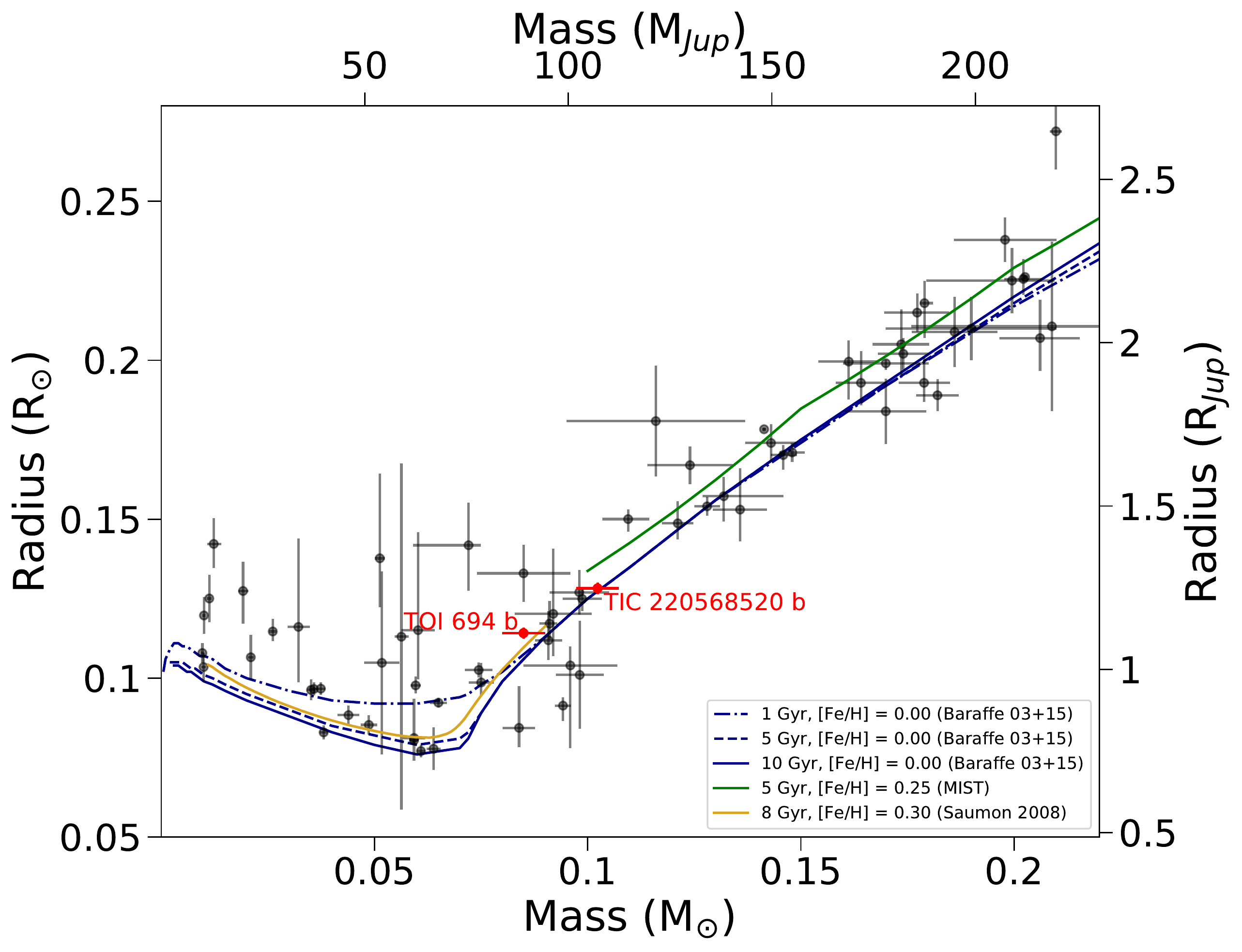}
\caption{Radius-mass diagram for the massive planets, brown dwarfs, and low mass stars listed in \tabr{comp_catalog}. The blue lines are models for low mass stars and substellar objects with solar metallicity from \citet{baraffe2015} and \citet{baraffe2003}. The green line is derived from MIST evolutionary tracks for [Fe/H] = 0.25. The gold line is a model from \citet{saumon2008_bd} for [Fe/H] = 0.3. RIK 72 b is a $\sim$ 60 \mjup\, transiting brown dwarf that is not shown due to its inflated radius of 3.10 \rjup\, \citep{david2019}.}
\label{fig:mass_rad}
\end{figure*}

\begin{figure*}
\includegraphics[width=\columnwidth]{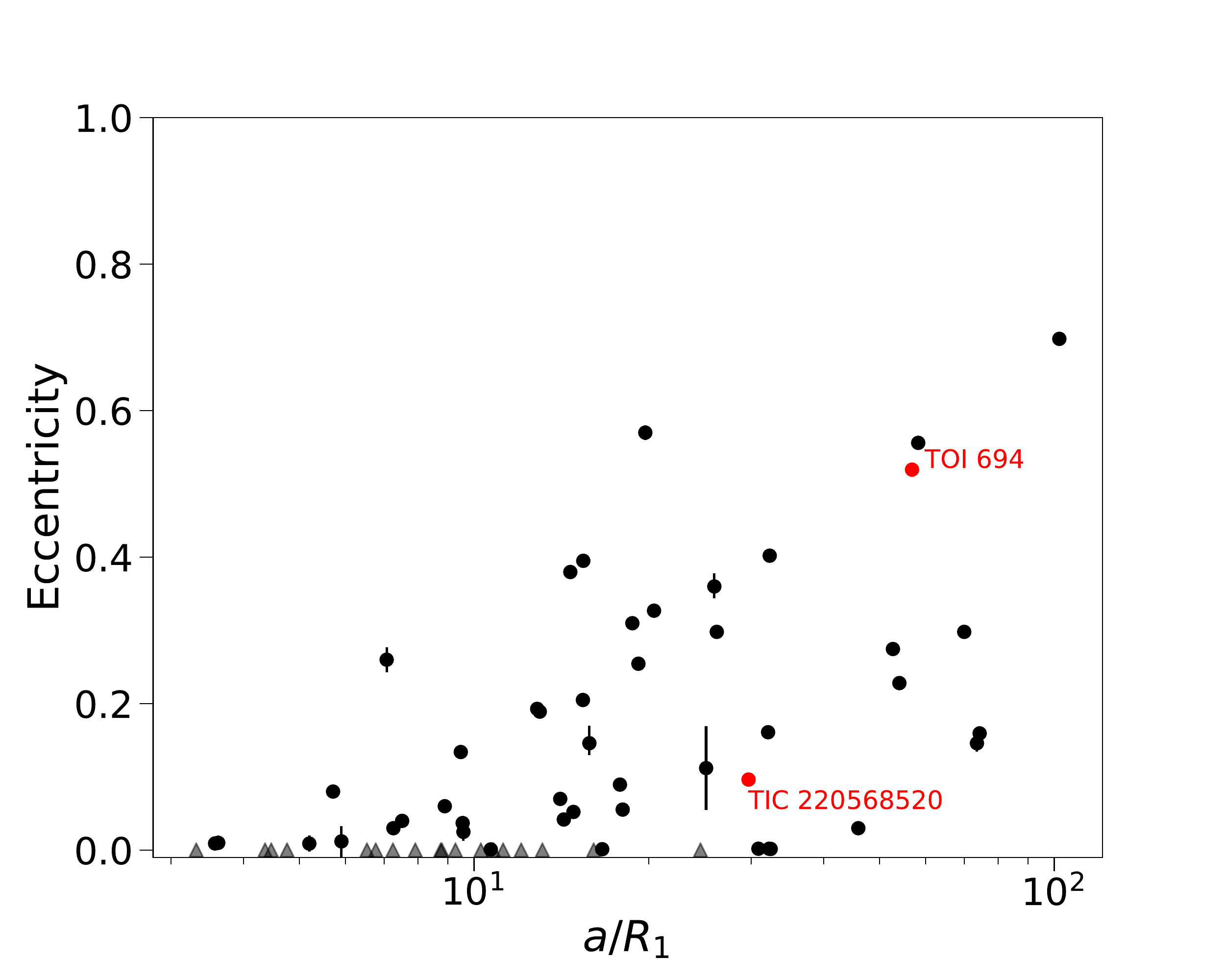}
\includegraphics[width=\columnwidth]{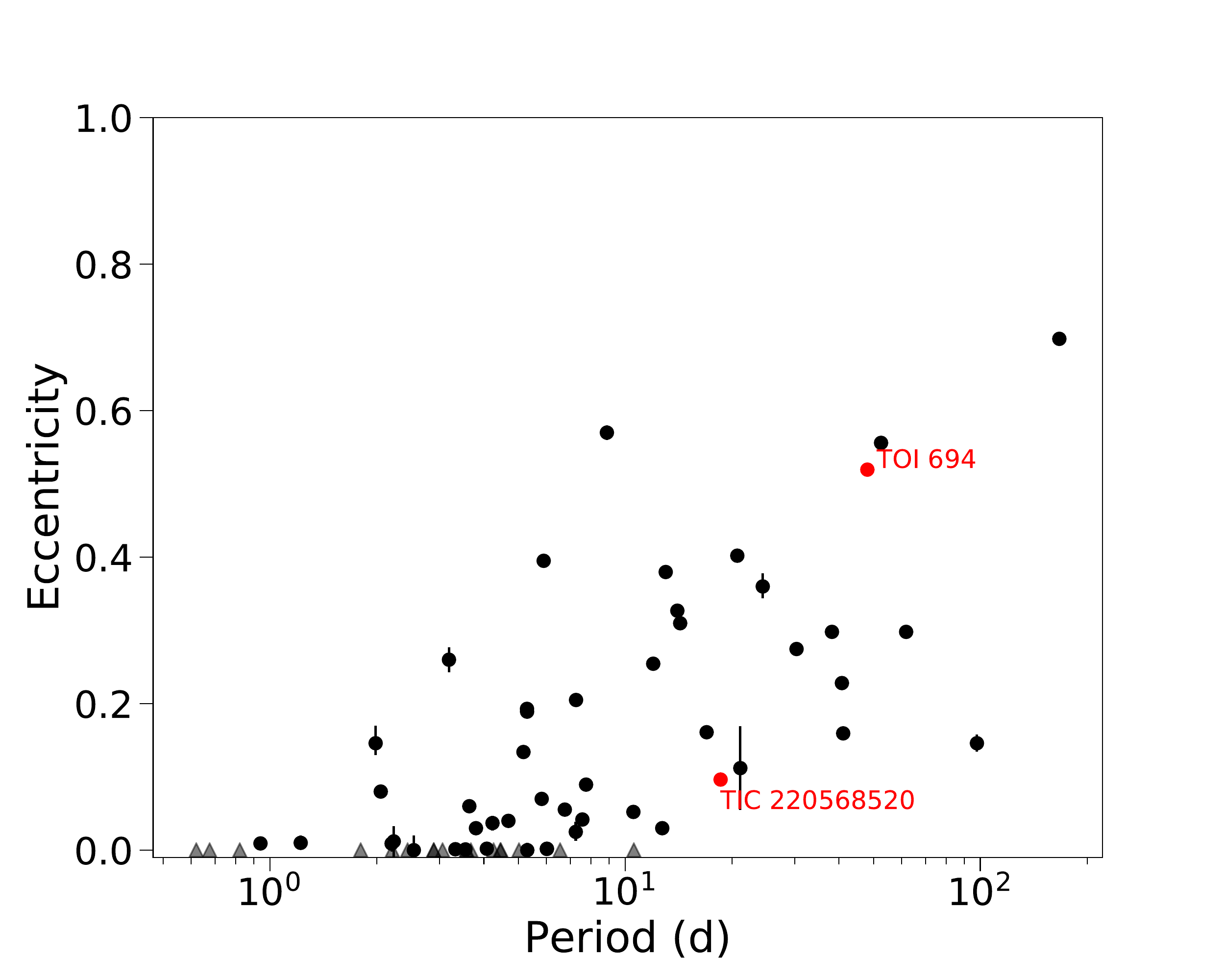}
\caption{Eccentricity as a function of scaled semi-major axis (left) and period (right) for the substellar and stellar companions listed in \tabr{comp_catalog}. The scaled semi-major axis for each system is determined using Kepler's Third Law. The two systems studied here are labeled and marked in red. Error bars are plotted for all systems, but for some they are smaller than the size of the marker. Faded triangles represent systems where the eccentricity was fixed to 0.}
\label{fig:per_ecc}
\end{figure*}

The two binary companions measured here are among the smallest stars with a measured radius and mass. To show them in the context of similar objects we plot in Figure~\ref{fig:mass_rad} the radius-mass diagram spanning brown dwarfs and small stars, with a mass range of 0.01~--~0.21~\msun. We mark on that diagram the new objects studied here (in red), and objects within that mass range that have a measured radius reported in the literature which we list in \tabr{comp_catalog}\footnote{The list given in \tabr{comp_catalog} is the result of our efforts to compile all objects reported in the literature within the mass range of 0.01 -- 0.21~\msun\ whose radius is also measured. While we recognize the possibility that a few objects may have unintentionally been omitted from such a compilation, such omission is highly unlikely to change the characaristics of the population of these objects, as presented in Figures~\ref{fig:mass_rad} and~\ref{fig:per_ecc} and discussed in the text.}, with the exception of the inflated brown dwarf RIK 72 b \citep{david2019}.


We also plot in the radius-mass diagram the theoretical isochrones for solar metallicity at ages of 1, 5, and 10 Gyr taken from \cite{baraffe2003, baraffe2015}, and the MESA isochrones and stellar tracks (MIST\footnote{\url{http://waps.cfa.harvard.edu/MIST/}}; \citealt{mist_dotter, mist_paxton11}) at 5 Gyr with solar metallicity and metallicity of [Fe/H] = 0.25. The latter is close to the metallicity we measure for the primary star in \target\ which we assume is also the metallicity of the secondary. 

The position in the radius-mass diagram of TOI~694 studied here is consistent with the 8 Gyr and [Fe/H] = 0.30 isochrone from \citet{saumon2008_bd}. In contrast, the position of \target\ is slightly {\it below} the MIST relation for [Fe/H] = 0.25, with a distance in radius of about 5\%. 
This small inconsistency is in the opposite direction of the inflated radius identified for stars of similar mass \citep[e.g.,][]{ribas06, Torres:2010, burrows2011, kesseli2018} and believed to be the result of enhanced magnetic fields in rapidly rotating stars \citep[e.g,][]{Chabrier2007}. However, the recent work of \cite{han2019} and \cite{vonBoetticher2019} showed that the radii of many low-mass fully convective stars are consistent with theoretical expectations without invoking enhanced magnetic fields. 

Fast rotation, which leads to enhanced magnetic fields, can result from spin-orbit tidal synchronization of short period systems \citep{mazeh2008}. The two systems studied here have relatively long orbital periods, of 18.6 days (TIC 220568520) and 48.1 days (TOI 694), longer than 95\% (TOI 694) and 87\% (TIC 220568520) of the systems listed in \tabr{comp_catalog}. Therefore these systems are not expected to have reached tidal synchronization. Even if the low-mass binary companions have reached pseudo-synchronization \citep{hut1981, zimmerman2017}, the rotation periods would be longer than the typical few days orbital periods of most of the systems in \tabr{comp_catalog}. Therefore, whether or not enhanced magnetic fields affect the radius of low-mass stars, it is not likely that they affect the radius of the two low-mass stars studied here.

Both systems studied here have a non-circular orbit, with well-measured orbital eccentricity. That is expected given their relatively long orbital periods, leading to a predicted orbital circularization time scale on the order of 10$^{14}$--10$^{15}$ years \citep{hilditch2001_timescales}. We show in \figr{per_ecc} the orbital eccentricity as a function of scaled semi-major axis and orbital period for all of the systems with brown dwarfs and low-mass stellar secondaries listed in \tabr{comp_catalog}. The figure shows TIC 220568520 with an eccentricity at the lower end of the eccentricity range of systems with similar orbital periods, and TOI 694 with an eccentricity close to the upper end of that of systems at similar orbital periods.

Tidal alignment for well separated binaries occurs on a similar timescale as tidal synchronization \citep{hut1981}. For the two systems studied here, the synchronization timescales are on the order of 10$^{12}$ years, meaning that these systems are not expected to be aligned due to tidal alignment mechanisms, although, they may have formed in aligned configurations. Therefore, measuring the primary star's obliquity, through the Rossiter-McLaughlin (RM) effect \citep[e.g.,][]{albrecht2007, gaudi2007, triaud2018_rm}, can probe their formation process \citep[e.g.,][]{pringle1989, bonnell92, bonnell94, fabrycky2007}. The predicted semi-amplitudes of the RM RV signal are 17 \ms\ and 35 \ms\ for TOI 694 and TIC 220568520, respectively \citep[Derived using Eq.~1 of][]{triaud2018_rm}.


The well-measured orbits and transits allow us to calculate the expected phase and duration of the secondary eclipses, where the low-mass star moves behind the primary star. A detection of a secondary eclipse and a measurement of its depth can constrain the temperature of the low-mass object, leading to a more complete characterization. The phase folded light curves of TOI 694 and TIC 220568520 during secondary eclipse phase are plotted in the right panels of Figures \ref{tess_lc_sec_694} and \ref{tess_lc_sec_220}, respectively. For each object we calculated the mean and standard deviation (root mean square) of all measurements determined to be within the secondary eclipse while removing 5-$\sigma$ outliers. For TOI 694 the measured secondary eclipse depth of the binned 2-minute cadence data is -75 $\pm$ 1164 ppm, and for TIC 220568520 it is 190 $\pm$ 811 ppm. While these are obviously non-detections, they are consistent with the expected \tess-band shallow secondary eclipses given the expected effective temperatures of the small low-mass companions. By assuming blackbody spectra and integrating across the \tess\ band we derive an expected secondary eclipse depth of $\approx$300~ppm for TOI 694 assuming its effective temperature is 2,600~K, and $\approx$580~ppm for TIC 220568520 assuming its effective temperature is 2,800~K.

While the secondary eclipses are not detected in \tess\ data and are too shallow to be detected from the ground in visible light, they are expected to be deeper, and perhaps detectable, in longer wavelengths. For example, in the $K$-band (centered around 2.1 \mic) the expected depths are $\approx$2300 ppm and $\approx$3900 ppm, for TOI 694 and TIC 220568520, respectively. More broadly, a comparison between the expected secondary eclipse depths and the uncertainty of the measured secondary eclipse depths (see previous paragraph) shows that for similar objects but with shorter orbital periods (and hence longer eclipses in phase) and brighter primary stars the noise level can potentially allow a detection of the secondary eclipse for low-mass stars.


\section{Summary}
\label{sec:sum}

We presented the discovery of two low-mass stars that are transiting (eclipsing) binary companions to Sun-like stars with eccentric orbits. The two low-mass stars are at the very bottom of the main sequence, close to and above the Hydrogen burning mass threshold. They join a short but growing list of low-mass stars with well-measured mass and radius. The relatively long orbital period of both systems means that the low-mass stars' radius is not likely to be impacted by enhanced magnetic fields, and we find their position in the radius-mass diagram to be consistent with theoretical models. 

These discoveries emphasize the potential of transit surveys for detecting rare transiting low-mass binary companions, since given the similarity in radius of objects between $\sim$1 Jupiter mass gas-giant planets through $\sim$100 Jupiter mass stars, transiting gas-giant planet candidates are also, by definition, low-mass star candidates. During the \tess\ Extended Mission the two targets studied here will be observed again, leading to a refined transit light curve measurement and tighter upper limits on the secondary eclipse or possibly a secondary eclipse detection.

\acknowledgments

We acknowledge the use of \tess\ Alert data. These data are derived from pipelines at the \tess\ Science Office and at the \tess\ Science Processing Operations Center.
Funding for the \tess\ mission is provided by NASA's Science Mission directorate.
This paper includes data collected by the \tess\ mission, which are publicly available from the Mikulski Archive for Space Telescopes (MAST).
Resources supporting this work were provided by the NASA High-End Computing (HEC) Program through the NASA Advanced Supercomputing (NAS) Division at Ames Research Center.
This research has made use of the NASA Exoplanet Archive, which is operated by the California Institute of Technology, under contract with NASA under the Exoplanet Exploration Program.
This work has been carried out within the framework of the National Centre of Competence in Research PlanetS supported by the Swiss National Science Foundation. 
MNG acknowledges support from MIT's Kavli Institute as a Torres postdoctoral fellow.
LAdS is supported by funding from the European Research Council (ERC) under the European Union’s Horizon 2020 research and innovation programme (project {\sc Four Aces}; grant agreement No 724427).
AJ acknowledges support from FONDECYT project 1171208 and by the Ministry for the Economy, Development, and Tourism's Programa Iniciativa Cient\'{i}fica Milenio through grant IC\,120009, awarded to the Millennium Institute of Astrophysics (MAS).

{\it Facilities:}
\facility{\tess}, 
\facility{ANU:2.3m (Echelle)}, 
\facility{Euler 1.2m (CORALIE)}
\facility{MPG/ESO:2.2m (FEROS)},
\facility{SMARTS:1.5m (CHIRON)}, 
\facility{SOAR (HRCam)}

\bibliographystyle{apj}
\bibliography{refs}


\clearpage

\begin{turnpage}
\begin{adjustwidth*}{-1.5cm}{-1.5cm}
\begin{deluxetable*}{lllllllllllllll}
\small
\tablecaption{Catalog of low mass companions with measured masses and radii}
\label{tab:toicatalog}
\tabletypesize{\scriptsize}
\tablehead{
\colhead{System} & \colhead{$M_2$} & \colhead{$\sigma_{M_2}$} & \colhead{$R_2$} & \colhead{$\sigma_{R_2}$} & \colhead{$P$} & \colhead{$e$} & \colhead{$\sigma_e$} & \colhead{$M_1$} & \colhead{$\sigma_{M_1}$} & \colhead{$R_1$} & \colhead{$\sigma_{R_1}$} & \colhead{$T_{eff}$} & \colhead{$\sigma_{T_{eff}}$} & \colhead{Reference} \\
\colhead{} & \colhead{($M_J$)} & \colhead{($M_J$)} & \colhead{($R_J$)} & \colhead{($R_J$)} & \colhead{(days)} & \colhead{} & \colhead{} & \colhead{($M_{\odot}$)} & \colhead{($M_{\odot}$)} & \colhead{($R_{\odot}$)} & \colhead{($R_{\odot}$)} & \colhead{($K$)} & \colhead{($K$)}
}

\startdata
Kepler-75 & 10.1 & 0.4 & 1.05 & 0.03 & 8.88 & 0.57 & 0.01 & 0.91 & 0.04 & 0.89 & 0.02 & 5200 & 100 & \citealt{bonomo2015} \\
CoRoT-27 & 10.4 & 0.6 & 1.01 & 0.04 & 3.57 & 0\tablenotemark{a} & - & 1.05 & 0.11 & 1.08 & $^{+0.16}_{-0.08}$ & 5900 & 120 & \citealt{parviainen2014} \\
WASP-18 & 10.52 & 0.38 & 1.165 & 0.057 & 0.94 & 0.0092 & 0.0028 & 1.281 & 0.069 & 1.230 & 0.047 & 6400 & 100 & \citealt{southworth2009} \\
XO-3 & 11.8 & 0.6 & 1.22 & 0.07 & 3.19 & 0.260 & 0.017 & 1.21 & 0.07 & 1.38 & 0.08 & 6429 & 100 & \citealt{winn2008} \\
HATS 70 & 12.9 & $^{+1.8}_{-1.6}$ & 1.384 & $^{+0.079}_{-0.074}$ & 1.888 & $<0.18$ & - & 1.78 & 0.12 & 1.881 & $^{+0.059}_{-0.066}$ & 7930 & $^{+630}_{-820}$ & \citealt{zhou2019} \\
Kepler-39 & 20.1 & $^{+1.3}_{-1.2}$ & 1.24 & $^{+0.09}_{-0.10}$ & 21.09 & 0.112 & 0.057 & 1.29 & $^{+0.06}_{-0.07}$ & 1.40 & 0.10 & 6350 & 100 & \citealt{bonomo2015} \\
CoRoT-3 & 22.0 & 0.7 & 1.037 & 0.069 & 4.26 & 0\tablenotemark{a} & - & 1.37 & 0.09 & 1.56 & 0.09 & 6740 & 140 & \citealt{deleuil2008} \\
KELT-1 & 27.38 & 0.93 & 1.116 & $^{+0.038}_{-0.029}$ & 1.22 & 0.01 & $^{+0.010}_{-0.007}$ & 1.335 & 0.063 & 1.471 & $^{+0.045}_{-0.035}$ & 6516 & 49 & \citealt{siverd2012} \\
NLTT 41135 & 33.7 & $^{+2.8}_{-2.6}$ & 1.13 & $^{+0.27}_{-0.17}$ & 2.889 & 0\tablenotemark{a} & - & 0.188 & $^{+0.026}_{-0.022}$ & 0.21 & $^{+0.016}_{-0.014}$ & 3230 & 130 & \citealt{irwin2010} \\
EPIC 219388192 & 36.8 & 1.0 & 0.937 & 0.032 & 5.29 & 0.1929 & 0.0019 & 1.01 & 0.04 & 1.01 & 0.03 & 5850 & 85 & \citealt{nowak2017} \\
WASP-128 & 37.5 & 0.8 & 0.94 & 0.02 & 2.21 & 0\tablenotemark{a} & - & 1.16 & 0.04 & 1.16 & 0.02 & 5950 & 50 & \citealt{hodzic2018} \\
CWW 89 A & 39.21 & 1.10 & 0.941 & 0.019 & 5.293 & 0.1891 & 0.0022 & 1.101 & $^{+0.039}_{-0.045}$ & 1.029 & 0.016 & 5755 & 49 & \citealt{carmichael2019} \\
KOI 205 & 39.9 & 1.0 & 0.807 & 0.022 & 11.72 & $<0.031$ & - & 0.925 & 0.033 & 0.841 & 0.020 & 5237 & 60 & \citealt{diaz2013} \\
TOI 1406 & 46.0 & 2.7 & 0.86 & 0.03 & 10.57415 & 0\tablenotemark{a} & - & 1.18 & 0.09 & 1.35 & 0.03 & 6290 & 100 & \citealt{carmichael2020} \\
EPIC 212036875 & 51 & 2 & 0.83 & 0.03 & 5.17 & 0.134 & 0.002 & 1.15 & 0.08 & 1.41 & 0.05 & 6230 & 90 & \citealt{persson2019} \\
TOI-503 & 53.7 & 1.2 & 1.34 & $^{+0.26}_{-0.15}$ & 3.677 & 0\tablenotemark{a} & - & 1.80 & 0.06 & 1.70 & 0.05 & 7650 & 160 & \citealt{subjak2019} \\
AD 3116 & 54.2 & 4.3 & 1.02 & 0.28 & 1.983 & 0.146 & $^{+0.024}_{-0.016}$ & 0.276 & 0.020 & 0.29 & 0.08 & 3184 & 29 & \citealt{gillen2017} \\
CoRoT-33 & 59 & $^{+1.8}_{-1.7}$ & 1.1 & 0.53 & 5.819 & 0.07 & 0.0016 & 0.86 & 0.04 & 0.94 & $^{+0.14}_{-0.08}$ & 5225 & 80 & \citealt{csizmadia2015} \\
RIK 72 & 59.2 & 6.8 & 3.10 & 0.31 & 97.76 & 0.146 & 0.0116 & 0.439 & 0.044 & 0.961 & 0.096 & 3349 & 142 & \citealt{david2019} \\
LHS 6343 C & 62.1 & 1.2 & 0.783 & 0.011 & 12.713 & 0.03 & 0.002 & 0.358 & 0.011 & 0.373 & 0.005 & 3431 & 21 & \citealt{montet2015} \\
KOI 415 & 62.1 & 2.7 & 0.790 & $^{+0.120}_{-0.070}$ & 166.8 & 0.698 & 0.002 & 0.94 & 0.06 & 1.250 & $^{+0.015}_{-0.010}$ & 5810 & 80 & \citealt{moutou2013} \\
WASP-30 & 62.50 & 1.20 & 0.951 & $^{+0.028}_{-0.024}$ & 4.16 & $<0.0044$ & - & 1.249 & $^{+0.032}_{-0.036}$ & 1.389 & $^{+0.033}_{-0.025}$ & 6202 & $^{+42}_{-51}$ & \citealt{triaud2013} \\
CoRoT-15 & 63.1 & 4.1 & 1.12 & $^{+0.30}_{-0.15}$ & 3.06 & 0\tablenotemark{a} & - & 1.32 & 0.12 & 1.46 & $^{+0.31}_{-0.14}$ & 6350 & 200 & \citealt{bouchy_corot_2011}b \\
TOI 569 & 63.8 & 1.0 & 0.75 & 0.02 & 6.55604 & 0\tablenotemark{a} & - & 1.21 & 0.03 & 1.48 & 0.03 & 5705 & 76 & \citealt{carmichael2020} \\
EPIC 201702477 & 66.9 & 1.7 & 0.757 & 0.065 & 40.74 & 0.2281 & 0.0026 & 0.870 & 0.031 & 0.901 & 0.057 & 5571 & 70 & \citealt{bayliss2017} \\
LP 261-75 & 68.1 & 2.1 & 0.898 & 0.015 & 1.882 & $<0.007$ & - & 0.300 & 0.015 & 0.313 & 0.005 & 3100 & 50 & \citealt{irwin2018} \\
NGTS-7A & 75.5 & $^{+3}_{-13.7}$ & 1.38 & $^{+0.13}_{-0.14}$ & 0.676 & 0\tablenotemark{a} & - & 0.480 & 0.075 & 0.61 & 0.06 & 3359 & $^{+106}_{-89}$ & \citealt{jackman2019} \\
KOI 189 & 78.0 & 3.4 & 0.998 & 0.023 & 30.36 & 0.2746 & 0.0037 & 0.764 & 0.051 & 0.733 & 0.017 & 4952 & 40 & \citealt{diaz2014} \\
Kepler-503 & 78.6 & 3.1 & 0.96 & $^{+0.06}_{-0.04}$ & 7.258 & 0.025 & $^{+0.014}_{-0.012}$ & 1.154 & $^{+0.047}_{-0.042}$ & 1.764 & $^{+0.08}_{-0.068}$ & 5690 & $^{+100}_{-110}$ & \citealt{canas2018} \\
EBLM J0555-57 & 87.90 & 3.98 & 0.821 & $^{+0.128}_{-0.058}$ & 7.758 & 0.0895 & $^{+0.0035}_{-0.0036}$ & 1.180 & $^{+0.082}_{-0.079}$ & 1.00 & $^{+0.14}_{-0.07}$ & 6386 & 124 & \citealt{vonBoetticher2019} \\
OGLE-TR-123 & 89.0 & 11.5 & 1.294 & 0.088 & 1.8 & 0\tablenotemark{a} & - & 1.29 & 0.26 & 1.55 & 0.1 & 6700 & 300 & \citealt{pont2006} \\
TOI 694 & 89.8 & 5.3 & 1.11 & 0.02 & 48.05125 & 0.5212 & 0.0021 & 0.967 & $^{+0.047}_{-0.04}$ & 0.998 & $^{+0.010}_{-0.012}$ & 5496 & $^{+87}_{-81}$ & This work \\
KOI-607 & 95.1 & $^{+3.4}_{-3.3}$ & 1.089 & $^{+0.089}_{-0.061}$ & 5.894 & 0.395 & $^{+0.0091}_{-0.0090}$ & 0.993 & $^{+0.050}_{-0.052}$ & 0.915 & $^{+0.031}_{-0.028}$ & 5418 & $^{+87}_{-85}$ & \citealt{carmichael2019} \\
J1219-39 & 95.4 & $^{+1.9}_{-2.5}$ & 1.140 & $^{+0.069}_{-0.049}$ & 6.76 & 0.05539 & $^{+0.00023}_{-0.00022}$ & 0.826 & $^{+0.032}_{-0.029}$ & 0.811 & $^{+0.038}_{-0.024}$ & 5412 & $^{+81}_{-65}$ & \citealt{triaud2013} \\
OGLE-TR-122 & 96.3 & 9.4 & 1.17 & $^{+0.20}_{-0.13}$ & 7.27 & 0.205 & 0.008 & 0.98 & 0.14 & 1.05 & $^{+0.20}_{-0.09}$ & 5700 & 300 & \citealt{pont_122_2005}a \\
K2-76 & 98.7 & 2.0 & 0.889 & $^{+0.025}_{-0.047}$ & 11.99 & 0.2545 & $^{+0.0070}_{-0.0065}$ & 0.964 & 0.026 & 1.171 & $^{+0.033}_{-0.060}$ & 5747 & $^{+64}_{-70}$ & \citealt{shporer2017} \\
C101186644 & 101 & 12 & 1.01 & $^{+0.06}_{-0.25}$ & 20.68 & 0.402 & - & 1.2 & 0.2 & 1.07 & 0.07 & 6090 & 200 & \citealt{talor2013} \\
J2343+29 & 103 & 7 & 1.236 & 0.068 & 16.95 & 0.161 & $^{+0.0015}_{-0.0027}$ & 0.864 & $^{+0.097}_{-0.098}$ & 0.854 & $^{+0.050}_{-0.060}$ & 5150 & $^{+90}_{-60}$ & \citealt{chaturvedi2016} \\
EBLM J0954-23 & 102.8 & $^{+6.0}_{-5.9}$ & 0.983 & 0.165 & 7.575 & 0.04186 & $^{+0.00094}_{-0.00092}$ & 1.166 & $^{+0.080}_{-0.082}$ & 1.23 & 0.17 & 6406 & 124 & \citealt{vonBoetticher2019} \\
KOI 686 & 103 & 5 & 1.22 & 0.04 & 52.51 & 0.556 & 0.0037 & 0.983 & 0.074 & 1.04 & 0.03 & 5750 & 120 & \citealt{diaz2014} \\
TIC 220568520 & 107.8 & 5.2 & 1.248 & 0.018 & 18.55741 & 0.0956 & $^{+0.0032}_{-0.0030}$ & 1.030 & $^{+0.043}_{-0.042}$ & 1.007 & $^{+0.010}_{-0.009}$ & 5589 & 81 & This work \\
HATS 550-016 & 115 & $^{+5}_{-6}$ & 1.46 & $^{+0.03}_{-0.04}$ & 2.05 & 0.08 & -1 & 0.97 & $^{+0.05}_{-0.06}$ & 1.22 & $^{+0.02}_{-0.03}$ & 6420 & 90 & \citealt{zhou2014}

\enddata
\label{tab:comp_catalog}
\end{deluxetable*}

\addtocounter{table}{-1}

\begin{deluxetable*}{lllllllllllllll}[!t]
\tablecaption{(Continued)}
\label{tab:toicatalog}
\tabletypesize{\scriptsize}
\tablehead{
\colhead{System} & \colhead{$M_2$} & \colhead{$\sigma_{M_2}$} & \colhead{$R_2$} & \colhead{$\sigma_{R_2}$} & \colhead{$P$} & \colhead{$e$} & \colhead{$\sigma_e$} & \colhead{$M_1$} & \colhead{$\sigma_{M_1}$} & \colhead{$R_1$} & \colhead{$\sigma_{R_1}$} & \colhead{$T_{eff}$} & \colhead{$\sigma_{T_{eff}}$} & \colhead{Reference} \\
\colhead{} & \colhead{($M_J$)} & \colhead{($M_J$)} & \colhead{($R_J$)} & \colhead{($R_J$)} & \colhead{(days)} & \colhead{} & \colhead{} & \colhead{($M_{\odot}$)} & \colhead{($M_{\odot}$)} & \colhead{($R_{\odot}$)} & \colhead{($R_{\odot}$)} & \colhead{($K$)} & \colhead{($K$)}
}

\startdata
OGLE-TR-106 & 121 & 22 & 1.76 & 0.17 & 2.54 & 0 & 0.02 & - & - & 1.31 & 0.09 & - & - & \citealt{pont2005}b \\
EBLM J1431-11 & 126.9 & $^{+3.77}_{-3.87}$ & 1.447 & $^{+0.0681}_{-0.0487}$ & 4.45 & 0\tablenotemark{a} & - & 1.200 & $^{+0.056}_{-0.055}$ & 1.114 & $^{+0.043}_{-0.028}$ & 6161 & 124 & \citealt{vonBoetticher2019} \\
HAT-TR-205-013 & 130 & 11 & 1.63 & 0.06 & 2.23 & 0.012 & 0.021 & 1.04 & 0.13 & 1.28 & 0.04 & 6295 & - & \citealt{beatty2007} \\
TIC 231005575 & 134.1 & 3.1 & 1.499 & 0.029 & 61.777 & 0.298 & $^{+0.004}_{-0.001}$ & 1.045 & 0.035 & 0.992 & 0.050 & 5500 & 85 & \citealt{gill2020a}a \\
HATS 551-021 & 138 & $^{+15}_{-5}$ & 1.53 & $^{+0.06}_{-0.08}$ & 3.64 & 0.06 & - & 1.10 & 0.10 & 1.20 & $^{+0.08}_{-0.01}$ & 6670 & 220 & \citealt{zhou2014} \\
EBLM J2017+02 & 142.2 & $^{+6.6}_{-6.7}$ & 1.489 & $^{+0.127}_{-0.097}$ & 0.822 & 0\tablenotemark{a} & - & 1.105 & $^{+0.074}_{-0.072}$ & 1.196 & $^{+0.080}_{-0.050}$ & 6161 & 124 & \citealt{vonBoetticher2019} \\
KIC 1571511 & 148.0 & 0.5 & 1.735 & $^{+0.005}_{-0.006}$ & 14.02 & 0.3269 & 0.0027 & 1.265 & $^{+0.036}_{-0.030}$ & 1.343 & $^{+0.012}_{-0.010}$ & 6195 & 50 & \citealt{ofir2012} \\
WTS 19g-4-02069 & 150 & 6 & 1.69 & 0.06 & 2.44 & 0\tablenotemark{a} & - & 0.53 & 0.02 & 0.51 & 0.01 & 3300 & 140 & \citealt{nefs2013} \\
K2-51 & 152.8 & $^{+3.4}_{-3.0}$ & 1.656 & $^{+0.031}_{-0.045}$ & 13.001 & 0.3797 & $^{+0.0090}_{-0.0058}$ & 1.068 & $^{+0.032}_{-0.029}$ & 1.695 & $^{+0.037}_{-0.049}$ & 5908 & $^{+63}_{-64}$ & \citealt{shporer2017} \\
TIC 238855958 & 155.0 & 3.1 & 1.664 & 0.029 & 38.195 & 0.298 &  & 1.514 & 0.037 & 2.159 & 0.037 & 6280 & 85 & \citealt{gill2020b}b \\
K2-67 & 168.9 & $^{+7.0}_{-7.5}$ & 1.942 & $^{+0.065}_{-0.116}$ & 24.388 & 0.36 & $^{+0.018}_{-0.016}$ & 0.916 & $^{+0.290}_{-0.031}$ & 1.399 & $^{+0.056}_{-0.079}$ & 5579 & $^{+78}_{-77}$ & \citealt{shporer2017} \\
EBLM J0543-56 & 171.9 & $^{+6.0}_{-6.2}$ & 1.877 & $^{+0.097}_{-0.068}$ & 4.464 & 0\tablenotemark{a} & - & 1.276 & $^{+0.072}_{-0.070}$ & 1.255 & $^{+0.054}_{-0.036}$ & 6223 & 124 & \citealt{vonBoetticher2019} \\
HATS 551-019 & 178 & 10 & 1.79 & 0.10 & 4.69 & 0.04 & - & 1.10 & $^{+0.05}_{-0.09}$ & 1.70 & 0.09 & 6380 & 170 & \citealt{zhou2014} \\
KIC 7605600 & 178 & 11 & 1.94 & $^{+0.01}_{-0.02}$ & 3.326 & 0.0013 & $^{+0.0043}_{-0.0008}$ & 0.53 & 0.02 & 0.501 & $^{+0.001}_{-0.002}$ & - & - & \citealt{han2019} \\
EBLM J1038-37 & 181.8 & $^{+6.9}_{-7.0}$ & 1.995 & $^{+0.107}_{-0.097}$ & 5.022 & 0\tablenotemark{a} & - & 1.176 & $^{+0.072}_{-0.070}$ & 1.132 & $^{+0.052}_{-0.048}$ & 5885 & 124 & \citealt{vonBoetticher2019} \\
EBLM J2349-32 & 182.3 & 6.3 & 1.966 & 0.049 & 3.5496972 & 0.001 & 0.002 & 0.991 & 0.049 & 0.965 & 0.022 & 6130 & 85 & \citealt{gill2019} \\
EBLM J1013+01 & 185.7 & $^{+7.9}_{-8.1}$ & 2.092 & 0.058 & 2.892 & 0\tablenotemark{a} & - & 1.036 & $^{+0.070}_{-0.072}$ & 1.036 & $^{+0.027}_{-0.026}$ & 5579 & 124 & \citealt{vonBoetticher2019} \\
EBLM J1115-36 & 187.4 & $^{+6.4}_{-6.2}$ & 1.877 & $^{+0.078}_{-0.058}$ & 10.543 & 0.0522 & $^{+0.0038}_{-0.0037}$ & 1.369 & 0.072 & 1.579 & $^{+0.048}_{-0.041}$ & 6605 & 124 & \citealt{vonBoetticher2019} \\
HATS 551-027 & 187.5 & $^{+2.1}_{-1.0}$ & 2.121 & $^{+0.068}_{-0.107}$ & 4.077 & 0.002 & - & 0.244 & 0.003 & 0.261 & $^{+0.006}_{-0.009}$ & 3190 & 100 & \citealt{zhou2015} \\
EBLM J2308-46 & 190.7 & 5.2 & 1.839 & 0.049 & 2.199187 & 0.009 & 0.011 & 1.223 & 0.049 & 1.534 & 0.041 & 6185 & 85 & \citealt{gill2019} \\
J0113+31 & 194.9 & 10.5 & 2.033 & 0.107 & 14.28 & 0.3098 & 0.0005 & 0.945 & 0.045 & 1.378 & 0.058 & 5961 & 54 & \citealt{gomez2014} \\
2MASS J0446+19 & 199 & 21 & 2.04 & 0.10 & 0.62 & 0\tablenotemark{a} & - & 0.47 & 0.05 & 0.56 & 0.02 & 3320 & 150 & \citealt{hebb2006} \\
T-Lyr-101662 & 207 & 13 & 2.32 & 0.07 & 4.23 & 0.037 & 0.01 & 0.77 & 0.08 & 1.14 & 0.03 & 6200 & 30 & \citealt{fernandez2009} \\
HATS 553-001 & 209 & $^{+11}_{-21}$ & 2.19 & 0.10 & 3.8 & 0.03 & - & 1.20 & 0.10 & 1.58 & $^{+0.08}_{-0.03}$ & 6230 & 250 & \citealt{zhou2014} \\
AD 3814 & 211.8 & 4.7 & 2.195 & $^{+0.061}_{-0.048}$ & 6.02 & 0.00194 & $^{+0.00253}_{-0.00057}$ & 0.3813 & 0.0074 & 0.3610 & 0.0033 & 3211 & $^{+54}_{-36}$ & \citealt{gillen2017} \\
Kepler-16 & 212.2 & 0.7 & 2.201 & 0.006 & 41.08 & 0.15944 & $^{+0.00062}_{-0.00061}$ & 0.6897 & $^{+0.0035}_{-0.0034}$ & 0.6489 & 0.0013 & 4450 & 150 & \citealt{doyle2011} \\
EBLM J0339+03 & 215.9 & $^{+9.7}_{-10.0}$ & 2.014 & $^{+0.117}_{-0.100}$ & 3.581 & 0\tablenotemark{a} & - & 1.036 & $^{+0.074}_{-0.076}$ & 1.210 & $^{+0.055}_{-0.052}$ & 6132 & 124 & \citealt{vonBoetticher2019} \\
OGLE-TR-125 & 219 & 35 & 2.05 & 0.26 & 5.3 & 0 & 0.01 & - & - & 1.94 & 0.18 & - & - & \citealt{pont2005}b \\
PTFEB132.707+19.810 & 219.8 & 1.5 & 2.647 & 0.117 & 6.016 & 0.0017 & 0.0006 & 0.3953 & 0.0020 & 0.363 & 0.008 & 3260 & 67 & \citealt{kraus2017}
\enddata
\tablenotetext{a}{Eccentricity fixed at 0}
\end{deluxetable*}
\end{adjustwidth*}
\end{turnpage}

\end{document}